\documentclass[lettersize,journal]{IEEEtran}
\usepackage{amsmath,amsfonts}
\usepackage{algorithm}       
\usepackage{algpseudocode}   

\newcounter{algsubstate}

\usepackage{array}
\usepackage[caption=false,font=normalsize,labelfont=sf,textfont=sf]{subfig}
\usepackage{textcomp}
\usepackage{stfloats}
\usepackage{url}
\usepackage{verbatim}
\usepackage{graphicx}
\def\BibTeX{{\rm B\kern-.05em{\sc i\kern-.025em b}\kern-.08em
    T\kern-.1667em\lower.7ex\hbox{E}\kern-.125emX}}
\usepackage{balance}

\usepackage{epsfig}
\usepackage{color,soul}
\usepackage[dvipsnames]{xcolor}
\RequirePackage{booktabs}
\usepackage{multirow}
\usepackage{caption}
\usepackage{makecell}
\usepackage{adjustbox}
\usepackage{rotating}
\usepackage{longtable}
\usepackage{tabularx}
\usepackage{lscape}

\begin{document}

\title{Partial-Reference IQA Based on Hermite-Gauss Structural Prediction and Texture Deviation}

\author{
	Paolo Giannitrapani, Elio D. Di Claudio, and Giovanni Jacovitti%
	\thanks{P. Giannitrapani is with the Department of Information Engineering, Electronics and Telecommunications (DIET), University of Rome ``Sapienza'', Rome, Italy (e-mail: paolo.giannitrapani@ieee.org). E. D. Di Claudio and G. Jacovitti were with the same department (e-mail: elio.diclaudio@uniroma1.it; gianni.iacovitti@gmail.com). EDICS No.: SMR-HPM.}}

\markboth{PROPOSAL2026}%
{Partial-Reference IQA Based on Hermite-Gauss Structural Prediction and Texture Deviation}

\maketitle

\begin{abstract}
We propose PreSPA (Partial-Reference Structural Prediction Approach), a Partial-Reference Image Quality Assessment framework that decomposes perceptual quality into two complementary indices. A structure-aware index, operating in a No-Reference manner, captures structural degradation through Hermite-Gauss prediction of the distorted gradient field and the angular variance of its curvature. A texture-sensitive index estimates local noise through a scalar prior $\mu$, obtained from energy differences between reference and distorted complex gradient maps on strong-edge regions and accumulated over weakly-structured ones, reflecting the perceptual leakage of degraded edges into surrounding textures. Crucially, $\mu$ is the only information extracted from the reference and is computed once per image pair, reducing the reference footprint to a single scalar value. The final score is produced by an affine fusion with only three interpretable parameters, making the method compact, transparent, and computationally efficient, with the viewing distance embedded into the operator scale and no dataset-specific calibration. Extensive evaluations on six standard benchmarks show that PreSPA consistently rivals or exceeds leading No-Reference approaches, while in several cases matching the accuracy of Full-Reference models.
\end{abstract}


\section{Introduction}

\IEEEPARstart{I}{mage} Quality Assessment (IQA) plays a central role in
virtually every stage of modern visual pipelines, from acquisition and
compression to transmission, restoration, and display. The reference quantity
in this field is the subjective response of human observers, conventionally
expressed as Mean Opinion Score (MOS) or Difference of Mean Opinion Score
(DMOS). Because collecting subjective ratings is impractical in real-time
or large-scale settings, the goal of objective IQA is to predict
MOS/DMOS algorithmically while remaining faithful to perceptual behavior
\cite{WANG04,SHEIKH06B,CHOW16}.

Objective IQA methods are grouped into three classes by the reference information they exploit: Full-Reference (FR) methods compute a discrepancy with the fully available pristine image; Reduced-/Partial-Reference (RR/PR) methods access only a compact reference summary; No-Reference (NR) methods operate solely on the distorted image. FR techniques achieve the highest accuracy but require the reference to be transmitted or stored with the test content. NR techniques are appealing for deployability, yet the most competitive ones rely on deep models with tens to hundreds of millions of parameters, large training corpora, and limited interpretability \cite{SU20,KE21,CHEN24,WANG23}. The PR setting sits in between but has received little attention: existing approaches extract pre-defined global features transmitted as side information \cite{WANG05C,SOUNDARARAJAN12}, leaving largely unexplored how a \emph{minimal} reference signal can guide an otherwise reference-free perceptual analysis.

A second limitation concerns how perceptual content is treated. Most metrics pool local measures into a single scalar without distinguishing the perceptual roles of different structures. Human observers, however, weigh locations unequally: strong isolated edges dominate scene interpretation, and the perceived integrity of surrounding texture depends on the fidelity of those edges, in line with Marr's hierarchical model of early vision \cite{MARR10}. The recent BELE estimator \cite{giannitrapani2025beleblurequivalentlinearized,GIANNITRAPANI25} showed that disentangling these roles yields interpretable, lightweight models with state-of-the-art accuracy, but entirely in the FR setting, relying on dense pixel-wise comparisons over both supports.

A third, often overlooked limitation concerns calibration. Objective scores are computed on the displayed image, whereas perception is governed by the retinal image, whose scale depends on the viewing distance. Most metrics either ignore this dependence or absorb it into a final calibration --- typically the five-parameter VQEG logistic rectification \cite{VQEG00} fitted per dataset --- which couples the metric to the specific subjective experiment and obscures the physical origin of the score. The FR estimator BELE \cite{giannitrapani2025beleblurequivalentlinearized,GIANNITRAPANI25} avoids this step by letting the viewing distance act directly on the scale of the feature-extraction operators. Whether the same principle can be carried over to a largely reference-free PR estimator has remained an open question.

In this work, we propose the \emph{Partial-Reference Structural Prediction Approach} (PreSPA), which exploits the perceptual asymmetry between edges and textures to minimize the reference footprint. The reference is used only to compute a \emph{single scalar prior} $\mu$, quantifying the energy deviation between reference and distorted complex gradient fields over structurally reliable regions. Computed once per image pair, $\mu$ is propagated to a fully No-Reference analysis built on two indices: a texture-fidelity index $Q_t$, a normalized signal-to-noise ratio modulated by $\mu$, and a structural index $Q_{\text{struct}}$, from a Hermite-Gauss self-prediction of the distorted gradient field. In both, the viewing distance enters by construction, rescaling the Gaussian kernels so that the indices are aligned on a common perceptual scale and combined through a low-order fusion, without dataset-dependent VQEG rectification. The estimator has three interpretable parameters --- the affine-fusion coefficients --- with two further constants ($a$, $\tau$) fixed a priori from metadata, and no learned weights.

The main contributions of this paper are summarized as follows:
\begin{itemize}
	\item{\emph{Minimal-reference design}: a single scalar, extracted from localized energy deviations on strong-edge regions, suffices to guide a perceptually grounded estimator, reducing the reference to this single value while preserving FR-like interpretability.}
	\item{\emph{Hermite-Gauss self-prediction}: a No-Reference structural index $Q_{\text{struct}}$ based on second-order Hermite-Gauss filters that locally regress the distorted gradient field, using the divergence and angular variance of the resulting tensor field as a reference-free descriptor of structural degradation.}
	\item{\emph{Cross-regional perceptual coupling}: the prior $\mu$ is evaluated on strong-edge regions ($\Omega_C$) but accumulated on weakly structured ones ($\Omega_H$), capturing how distortions in dominant structures leak into surrounding textures, consistently with Marr's raw primal sketch.}
	\item{\emph{Training-free estimator with affine-only alignment}: the viewing distance is embedded into the operator scale rather than into a post-hoc calibration, reducing the five-parameter VQEG logistic rectification to a plain three-coefficient affine alignment. PreSPA is, to our knowledge, the first PR method to adopt this principle, recently introduced for the FR estimator BELE.}
	\item{\emph{Comprehensive validation}: on six benchmarks (LIVE DBR2, TID2013, CSIQ, LIVE MD, KADID-10K, PIPAL), PreSPA consistently rivals or surpasses leading NR deep models, while matching state-of-the-art FR accuracy on several distortion classes.}
\end{itemize}

The remainder of the paper is organized as follows. Sec.~\ref{sec:Related Works} reviews the FR, NR, and PR/RR-IQA literature; Sec.~\ref{sec:Background} recalls the underlying theoretical framework; Sec.~\ref{sec:Proposed Method} introduces the PreSPA framework, its texture and structural indices, and the fusion stage; Sec.~\ref{sec:Experiments} reports the experimental evaluation; and Sec.~\ref{sec:Conclusion} concludes the paper.

\section{Related Works}
\label{sec:Related Works}

\noindent We organize the IQA literature according to the quality-prediction mechanism, with explicit attention to the reference-access protocol (FR, RR/PR, NR) along which PreSPA positions itself. We distinguish three families: \emph{perceptually motivated} models, which incorporate Human Visual System (HVS) mechanisms; \emph{classical signal-fidelity and hybrid analytic} methods, combining handcrafted features with structural or statistical measures; and \emph{data-driven} models, from early CNNs to transformer-CNN hybrids and large multimodal models. A dedicated subsection then covers the under-explored Reduced/Partial-Reference setting that is the natural home of PreSPA.

\subsection{Classification of IQA Approaches}
\label{subsec:Classification of IQA Approaches}

\noindent \emph{Perceptually motivated} models quantify differences between reference and distorted images by emulating early-stage HVS mechanisms, focusing on low-level features such as contrast sensitivity, spatial frequency response, and local structure \cite{WINKLER99,CHANDLER07}. They generally assume a relatively uniform perceptual processing across observers, neglecting higher-level cognitive factors --- such as emotion, education, or past experience --- that also influence subjective quality judgments \cite{CHANDLER06,VU08,CAVANAGH11}. Widely adopted FR methods belonging to this family, such as FSIM \cite{ZHANG11} and GMSD \cite{XUE14}, extract features inspired by contrast sensitivity, phase congruency, and gradient magnitude.

A complementary class is that of \emph{cognitive models}, which interpret degradation as visual information loss \cite{MARR10,LARSON10}: PSNR is the simplest instance, while VIF measures the loss of Shannon mutual information between reference and distorted images \cite{SHEIKH06}. Two information-theoretic mechanisms can be distinguished: the \emph{Fisher-information} one underlies BELE \cite{giannitrapani2025beleblurequivalentlinearized,GIANNITRAPANI25}, which disentangles strong-edge and texture degradation via Positional Fisher Information and a Complex PSNR term at state-of-the-art FR accuracy with three parameters; the \emph{Shannon channel-capacity} one models the distorted image as input and output of a noisy \emph{self-channel}, quantifying the retained information as a normalised capacity. PreSPA belongs to this second sub-current, inheriting BELE's edge--texture decomposition while replacing the full-reference Fisher machinery with a self-predictive formulation needing only a minimal scalar prior. Classical metrics such as SSIM \cite{WANG04} and MS-SSIM \cite{WANG03} blend psycho-physical and cognitive principles via multi-scale pooling.

A second family consists of \emph{classical signal-fidelity and hybrid analytic methods}, combining handcrafted features with structural or statistical measures. 2stepQA \cite{YU19} fuses no-reference (NIQE) and full-reference (MS-SSIM) assessments multiplicatively, while RVSIM \cite{YANG18} employs Log-Gabor filters and monogenic-signal analysis weighted by contrast sensitivity. Saliency-based variants such as SG-ESSIM \cite{VARGA22} and SWLGV \cite{VARGA22B} further incorporate visual attention into the SSIM framework. These methods retain interpretability and a parameter count in the order of units, but rely on the full reference image.

A third, increasingly dominant family is that of \emph{data-driven models}, which learn perceptual features from data. Within FR-IQA, CNN-based methods include TOPIQ-FR \cite{CHEN24}, a cross-feature attention network (with a PIPAL-retrained variant \cite{PIPAL20}), DISTS \cite{DING22}, balancing texture and structure similarity in feature space, LPIPS \cite{ZHANG18}, based on AlexNet or VGG features, PIEAPP \cite{PRASHNANI18}, and the semantic-guided SG-JND \cite{CAO24}. Hybrid transformer-CNN models include IQT \cite{CHEON21}, CQT \cite{OH22}, and AHIQ \cite{LAO22}, which combine transformers with CNN or deformable-convolution components for texture-rich and GAN-generated content. Finally, multimodal large pre-trained models such as CLIPIQA and CLIPIQA+ \cite{WANG23} exploit CLIP embeddings, while Q-Align \cite{WU24QALIGN} and Compare2Score \cite{ZHU24COMPARE2SCORE} reformulate IQA as a discrete-rating or pairwise-preference task solved by large multimodal models.

In the No-Reference setting, early statistical approaches --- DIIVINE \cite{MOORTHY11DIIVINE}, BLIINDS-II \cite{SAAD12}, BRISQUE \cite{MITTAL12}, NIQE \cite{MITTAL13B} --- extract Natural Scene Statistics (NSS) for a regressor. CNN-based methods (CNNIQA \cite{KANG14CNNIQA}, HyperIQA \cite{SU20}, MUSIQ \cite{KE21}, MANIQA \cite{YANG22MANIQA}, TOPIQ-NR \cite{CHEN24}) set the current state of the art, while self-supervised methods such as Re-IQA \cite{SAHA23REIQA} and ARNIQA \cite{AGNOLUCCI24ARNIQA} reduce reliance on labeled data. As in the FR case, these methods share large parameter counts and limited interpretability \cite{WANG19}.

\subsection{Reduced and Partial-Reference IQA}
\label{subsec:Reduced and Partial-Reference IQA}

\noindent Reduced-Reference (RR) and Partial-Reference (PR) IQA has received markedly less attention, despite its relevance when transmitting the full reference is infeasible (broadcasting, network monitoring, restoration diagnostics). The seminal wavelet-domain model of Wang and Simoncelli \cite{WANG05C} transmits a small set of reference features as side information; RRED \cite{SOUNDARARAJAN12} formalises this via entropic differencing, and more recent models exploit free-energy principles \cite{GU13B} or shallow learned features. In all cases the reference summary is a vector of pre-defined statistics whose dimensionality, though reduced, is non-trivial and fixed by design.

In all these approaches the side information enters symmetrically --- the same statistics are computed from both images and compared --- mirroring the FR philosophy of feature matching rather than exploiting an asymmetry between reference-derived priors and reference-free analysis. To the best of our knowledge, no prior PR-IQA work reduces the reference footprint to a \emph{single scalar} computed only on perceptually privileged regions, used to modulate a subsequent fully No-Reference analysis, as PreSPA does.

\subsection{Limitations of Existing Methods}
\label{subsec:Limitations of Existing Methods}

\noindent Most FR-IQA methods exhibit unequal sensitivity to different impairments \cite{SHEIKH06B,CAPODIFERRO12}, only partially mitigated by multi-resolution aggregation \cite{WANG03,XUE14}. A second, more fundamental limitation is the dependence on viewing distance, outlined in Sec.~I: most methods lack an explicit model of this effect, addressing it only implicitly through multi-scale pooling or, in some cases, by rescaling images \cite{GU15} or learning distance as a parameter \cite{BOSSE18}. A more principled alternative, introduced for FR-IQA in BELE \cite{giannitrapani2025beleblurequivalentlinearized,GIANNITRAPANI25}, embeds the viewing distance directly into the operator scale; transferring this principle to a largely reference-free estimator is one of the questions addressed here.

Deep models add further limitations: parameter counts of millions to hundreds of millions \cite{ZHANG18,KE21,CHEN24} entail high cost, limited interpretability, and overfitting risk, while generalisation is imperfect --- models trained on one benchmark (e.g., KADID-10K) often degrade on others (e.g., PIPAL) and require dataset-specific fine-tuning. Multimodal models such as CLIPIQA and Q-Align further inherit semantic biases from their pre-training corpora that do not always align with low-level quality.

In the RR and PR setting, the reference summary is fixed by design and computed globally, with no perceptual guidance on where the most informative content lies; moreover, the scarcity of recent work means PR-IQA has not benefited from the structural insights --- edge--texture decomposition, viewing-distance modelling, complex-domain analysis --- that have driven recent FR-IQA advances. PreSPA fills this gap, recasting PR-IQA as a minimal-reference, perceptually-grounded modulation of an otherwise fully No-Reference analysis.

\section{Background}
\label{sec:Background}

\begin{figure*}[!t]
	\centering
	\includegraphics[width=6.0in]{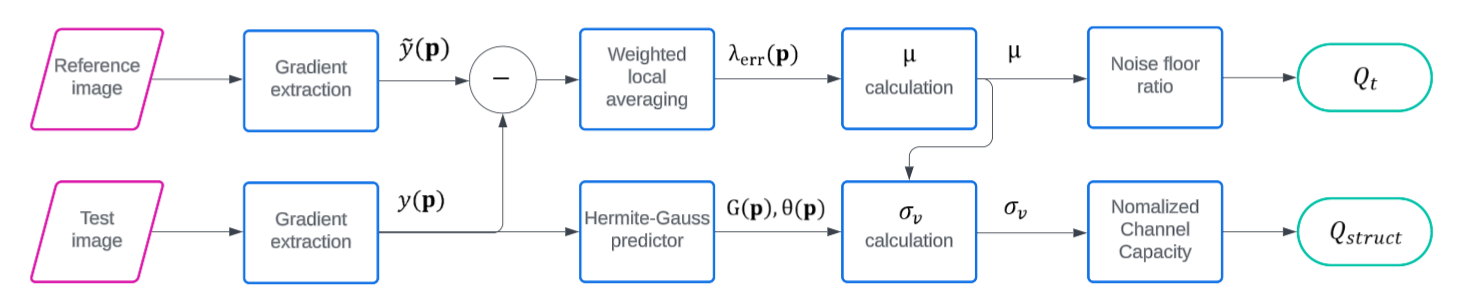}
	\caption{Flowchart illustrating the extended computational pipeline for deriving the texture index $Q_t$ and the structural information capacity $Q_{struct}$. The process starts with gradient extraction from both the test and reference images. A weighted local averaging is applied to compute the smoothed gradient energy for the residual map ($\lambda_{err}(\textbf{p})$), while the Hermite-Gauss predictor is applied only to the distorted image's complex map. The modulation index $\mu$, derived from $\lambda_{err}(\textbf{p})$, is used both to compute the texture index $Q_t$ and to estimate $\sigma_v$, which is required by the Normalized Channel Capacity block to obtain $Q_{struct}$.}
	\label{fig:PreSPA_model}
\end{figure*}

\noindent We briefly recall the Virtual Receptive Field model introduced in \cite{GIANNITRAPANI25, giannitrapani2025beleblurequivalentlinearized}, as a foundation for defining the concept of visual map and the smoothed gradient energy of the detail, both of which are essential to the proposed method.

The \emph{retinal model} targets the foveal region, where light receptors are distributed over Cartesian coordinates $\mathbf{p} \equiv (x_1, x_2)$. Image projection onto the retina yields a luminance signal $I(\mathbf{p})$, with chrominance ignored. A receptor density of approximately 60 samples per visual degree is assumed. Due to their non-uniform placement, the generalized sampling theory from \cite{LANDAU67} ensures accurate representation of $I(\mathbf{p})$, provided its 2D spatial spectrum remains within $(-30, +30)$ cycles/degree (or $(-1/2, +1/2)$ cycles/arcmin), matching the Snellen acuity limit.

The \emph{display model} assumes a regular pixel grid with resolution $R$ (pixels/mm), yielding a pixel pitch $d = 1/R$. To align display sampling with retinal resolution, a viewing distance $\delta_0$ is required such that $\delta_0 \times \tan(1) = 1/R = d$, where angles are in arcminutes and lengths in millimeters. This \emph{nominal viewing distance} $\delta_0$ ensures a perceptually lossless, non-redundant mapping from the display to the retina.

When the viewing distance differs from the nominal, i.e., at a generic distance $\delta$, the spectral content of the retinal image scales proportionally, resulting in a spatial frequency band of $\displaystyle\left(-\delta/2\delta_0, +\delta/2\delta_0\right)$ $cycles/arcmin$ or $\displaystyle\left(-30\;\delta/\delta_0, +30\;\delta/\delta_0\right)$ $cycles/degree$ in both horizontal and vertical directions, as prescribed by the Nyquist criterion. Consequently, if $\delta < \delta_0$, the bandwidth is reduced, implying a loss of perceived detail. Conversely, if $\delta > \delta_0$, the bandwidth increases, potentially enhancing perceived sharpness.

Within this framework, the visual response of the retina is modeled by receptive fields that compute a spatially weighted sum of the luminance component $I(\mathbf{p})$ in the neighborhood of each point $\mathbf{p}$. This is formalized via a spatial convolution between $I(\mathbf{p})$ and the so-called Virtual Receptive Field (VRF) $h(\mathbf{p})$:
\begin{equation}
	\begin{array}{l}
		y\left(\mathbf{p}\right)=I\left(\mathbf{p}\right)\ast h\left(\mathbf{p}\right) \vspace{2mm} \\
		h\left(\mathbf{p}\right)=Re{\left\{h\left(\mathbf{p}\right)\right\}}+jIm{\left\{h\left(\mathbf{p}\right)\right\}}
	\end{array}
	\label{eqn:visualmap}
\end{equation}
where $*$ denotes convolution. The output $y(\mathbf{p})$ is referred to as the \emph{visual map}, while $h(\mathbf{p})$ serves as a complex-valued Point Spread Function, i.e., the visual response to an impulse stimulus in a dark field \cite{DICLAUDIO21}.

The VRF model is specifically designed to extract the Positional Fisher Information, which quantifies the localizability of features in the image and supports gradient-based analysis \cite{GIANNITRAPANI25}.

Expressed in polar coordinates — $r = \sqrt{x_1^2 + x_2^2}$ and $\varphi = \tan^{-1}(x_2 / x_1)$ — the VRF adopts a polar-separable form:
\begin{equation}
	h(r,\varphi)=\frac{r}{2\pi s_G^2}e^{-r^2/(2s_G^2)}e^{j\varphi},
	\label{eqn:hmodel}
\end{equation}
where $s_G$ is the spread parameter of the VRF, controlling the extent of spatial integration.

This VRF is an eigenfunction of the two-dimensional Fourier transform \cite{DICLAUDIO10}, whose transform yields the \emph{Virtual Neural Transfer Function} (VNTF):
\begin{equation}
	H(\rho,\vartheta) = j 2\pi\rho e^{j\vartheta} e^{-s_G^2 \rho^2},
\end{equation}
where $\rho = \sqrt{f_1^2 + f_2^2}$ and $\vartheta = \tan^{-1}(f_2 / f_1)$ denote the radial and angular frequency coordinates.

Empirical studies in \cite{CAMPBELL65} confirm that this VNTF closely models the Contrast Sensitivity Function of the human visual system, as further discussed in \cite{MANNOS74}.

The model incorporates key aspects of human vision, such as \emph{orientation selectivity} and \emph{spatial frequency selectivity}, in accordance with the foundational principles established in \cite{DAUGMAN83}.

The visual map of a pristine (non-degraded) image on the display ${\widetilde{I}}_D\left(\mathbf{p}\right)$ projected onto the retina is:
\begin{equation}
	\widetilde{y}\left(\mathbf{p}\right)={\widetilde{I}}_D\left(\mathbf{p}\frac{\delta}{\delta_0}\right)\ast h(\mathbf{p})
\end{equation}
and
\begin{equation}
	\widetilde{\lambda}\left(\mathbf{p}\right)=\sum_{\mathbf{q}}{w_\mathbf{p}(\mathbf{q})^2\left|\widetilde{y}\left(\mathbf{p}-\mathbf{q}\right)\right|^2}
\end{equation}
is the energy of the detail of a pristine image, extracted by a window $w_\mathbf{p}(\mathbf{q})$, centered on $\mathbf{p}$ \cite{NERI04}.

Likewise, for the same detail of a degraded version of the image:
\begin{equation}
	y\left(\mathbf{p}\right)={\widetilde{I}}_D\left(\mathbf{p}\frac{\delta}{\delta_0}\right)\ast b\left(\mathbf{p}\frac{\delta}{\delta_0}\right)\ast h(\mathbf{p})
\end{equation}
is the visual map of a blurred version of a pristine image on the display ${\widetilde{I}}_D\left(\mathbf{p}\right)\ast b\left(\mathbf{p}\right)$ projected onto the retina ($b\left(\mathbf{p}\right)$ is the contribution of the Gaussian blur) and
\begin{equation}
	\lambda\left(\mathbf{p}\right)=\sum_{\mathbf{q}}{w_\mathbf{p}(\mathbf{q})^2\left|y\left(\mathbf{p}-\mathbf{q}\right)\right|^2}
\end{equation}
is the energy of the detail of a degraded image.

The smoothed gradient energy of the detail, computed on the reference and on the distorted image, constitutes the elementary building block of the proposed estimator. In the following section, these quantities are combined in two complementary ways: a Partial-Reference texture index, which compares reference and distorted energies only to extract a single scalar prior, and a No-Reference structural index, which predicts the expected gradient structure from the distorted image alone. 

\section{Proposed Method}
\label{sec:Proposed Method}

\noindent This section presents the proposed Partial-Reference Image Quality Assessment framework, which combines a texture-aware PR index and a structure-sensitive NR index. The overall design is motivated by the observation that texture degradation and geometric distortions often manifest independently, yet both significantly impact perceived image quality. By decoupling their estimation, the method ensures higher robustness and interpretability.

The proposed system relies on two complementary processing branches:
\begin{itemize}
	\item{A Partial-Reference (PR) branch that estimates a scalar noise prior $\mu$ from localized energy deviations between distorted and reference gradient maps, enabling perceptual texture degradation estimation.}
	\item{A No-Reference (NR) branch that evaluates structural inconsistencies such as edge deformation and curvature noise, using only the distorted image and the noise prior $\mu$.}
\end{itemize}

To ensure interpretability and facilitate implementation, each component is based on simple, spatial-domain operations without reliance on deep learning or pre-trained features.

\subsection{Overview and Processing Pipeline}
\label{subsec:Overview and Processing Pipeline}

Fig. \ref{fig:PreSPA_model} illustrates the processing pipeline, which combines a minimal Partial-Reference component with a fully No-Reference structural analysis:
\begin{itemize}
	\item{\emph{Gradient extraction}: gradient magnitude maps are computed from both reference and distorted images to characterize local texture energy.}
	\item{\emph{Noise prior $\mu$} (the only PR component): a scalar prior is derived from the energy error between predicted and observed gradients, restricted to strong-edge regions $\Omega_C$ but accumulated over weakly structured areas $\Omega_H$, capturing the perceptual propagation of structural distortions into texture zones. It is computed once and used as a fixed prior.}
	\item{\emph{Texture index $Q_t$}: a saturating signal-to-noise score relative to a perceptual threshold, in which the signal strength $\sigma_s^2$ is modulated by $\mu$.}
	\item{\emph{Hermite-Gauss structural prediction}: second-order Hermite-Gauss filters $(2,0)$ and $(0,2)$, fitted by weighted regression over a Gaussian window, yield a coefficient field $B(x,y)$ encoding local curvature and orientation.}
	\item{\emph{Structure-adaptive noise $\sigma_v^2$}: divergence and angular variance of $B(x,y)$ form a structural instability measure that perceptually modulates $\mu$, suppressing unstable regions.}
	\item{\emph{Structural index $Q_{\text{struct}}$}: a reference-free, rate-distortion-inspired signal-to-noise score driven by the predictive gain, signal energy, and modulated noise.}
	\item{\emph{Fusion}: the PR and NR components are combined through a low-order affine mapping.}
\end{itemize}

\subsection{Partial-Reference Texture Index}
\label{subsec:Partial-Reference Texture Index}

\begin{figure*}[!t]
	\centering
	\includegraphics[width=1.3in]{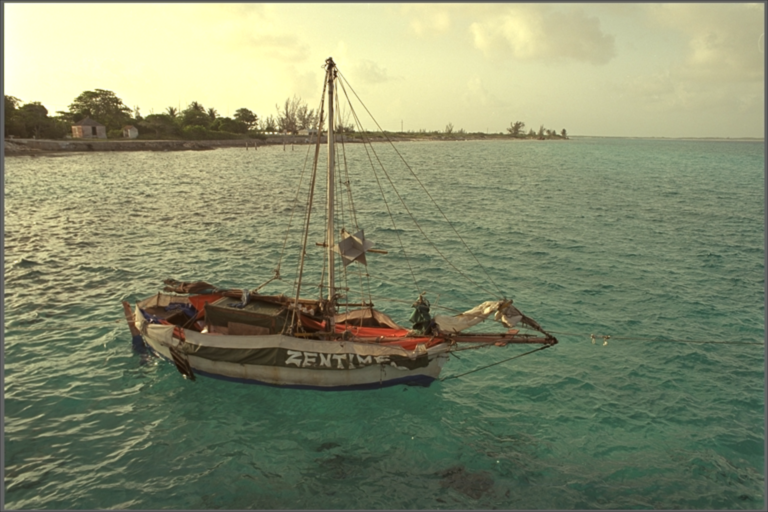}
	\includegraphics[width=1.3in]{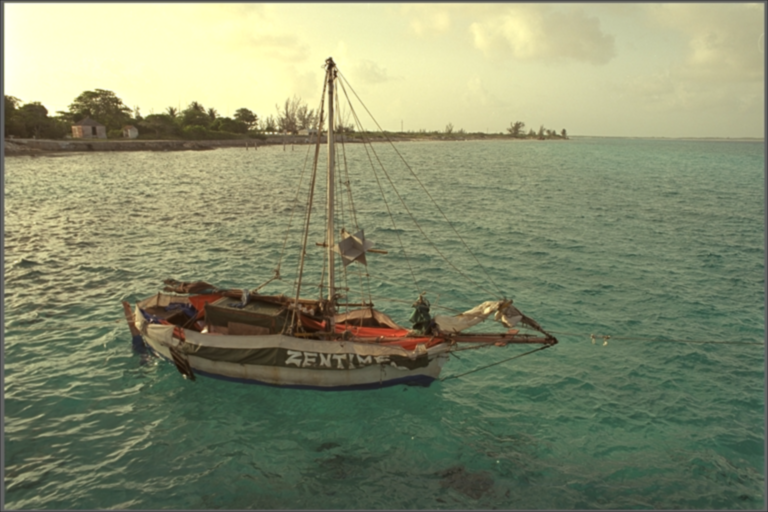}
	\includegraphics[width=1.3in]{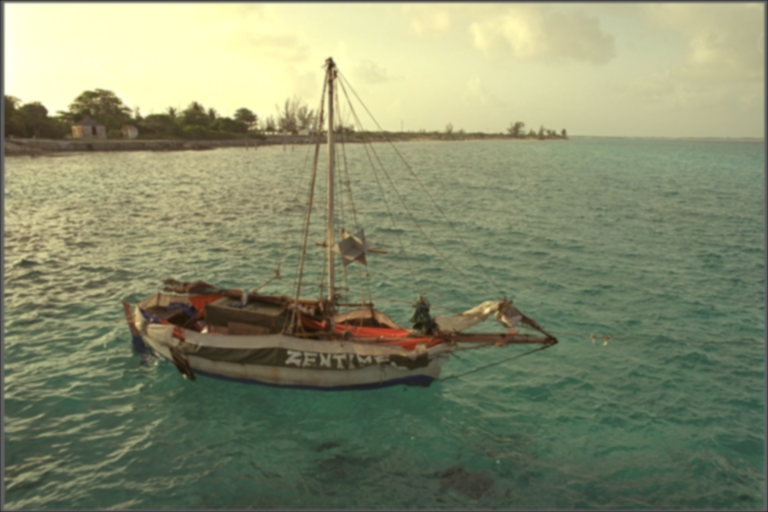}
	\includegraphics[width=1.3in]{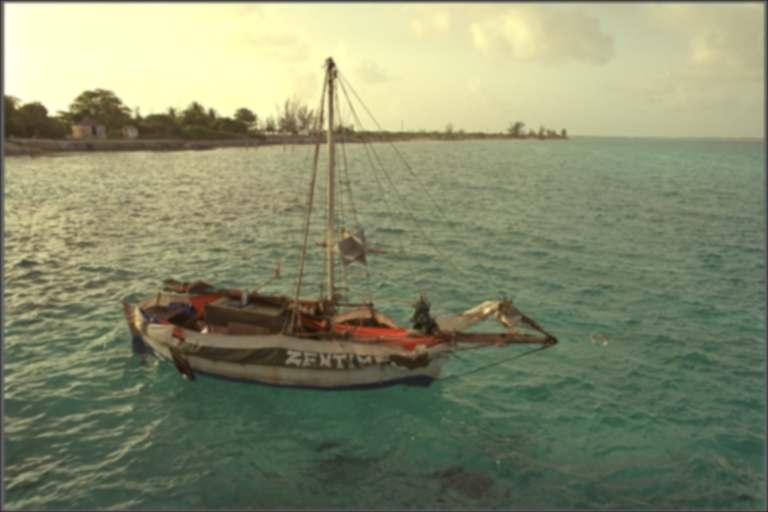}
	\includegraphics[width=1.3in]{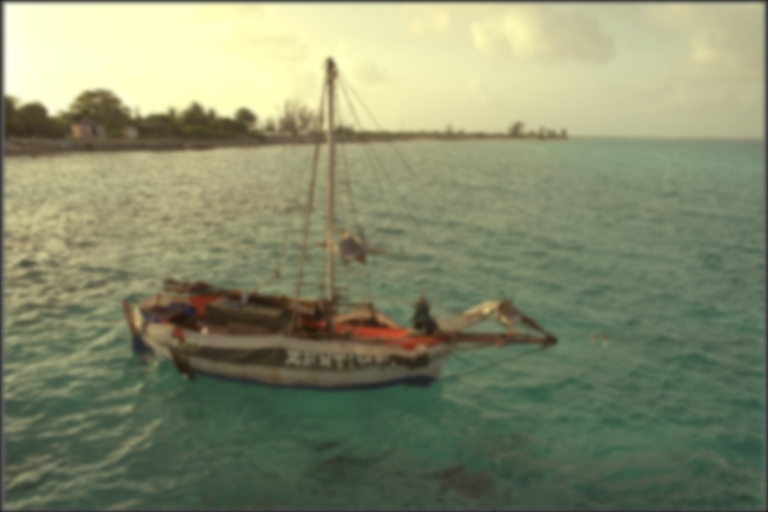} \\
	\vspace{1.5mm}
	\includegraphics[width=1.3in]{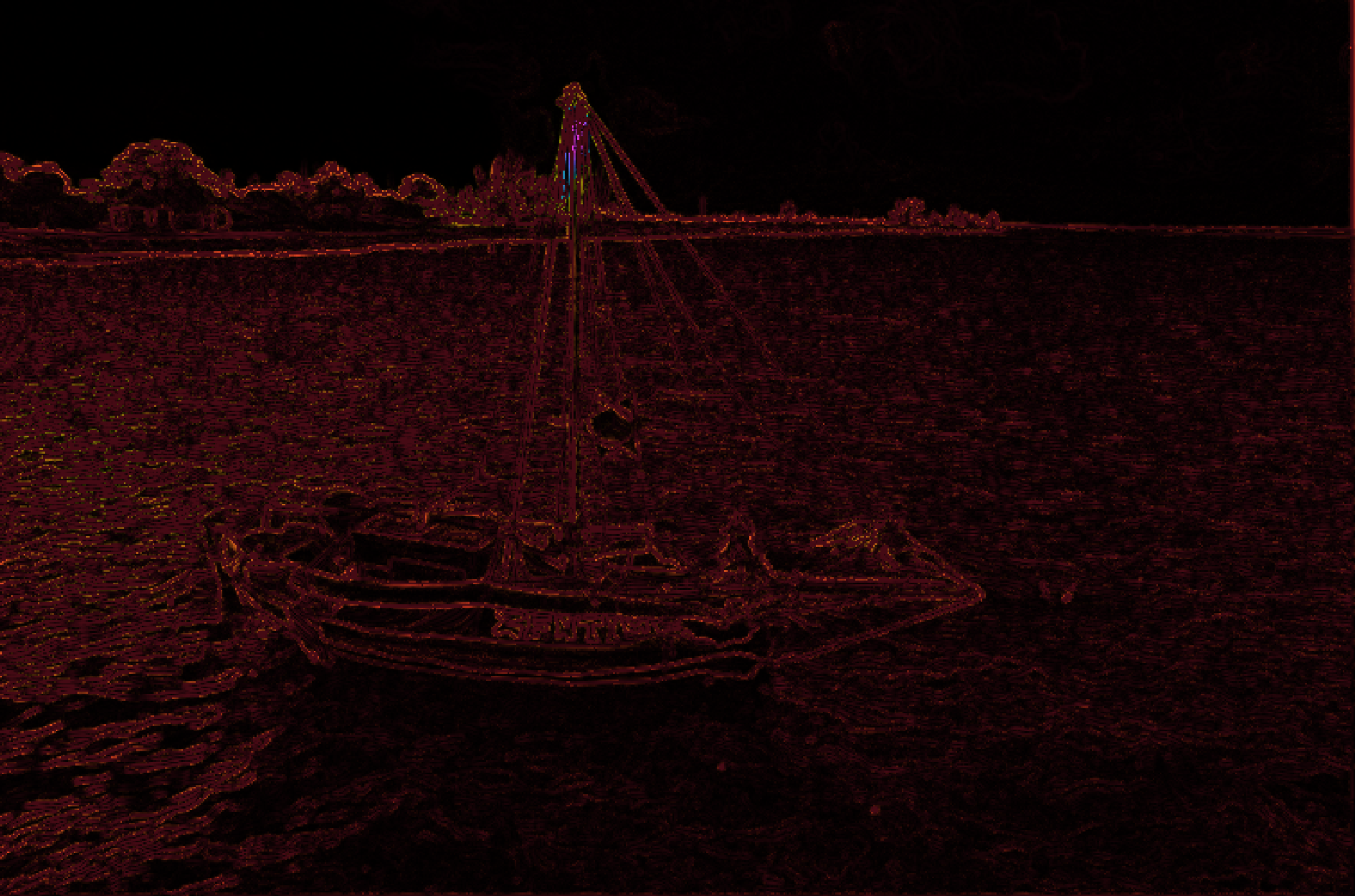}
	\includegraphics[width=1.3in]{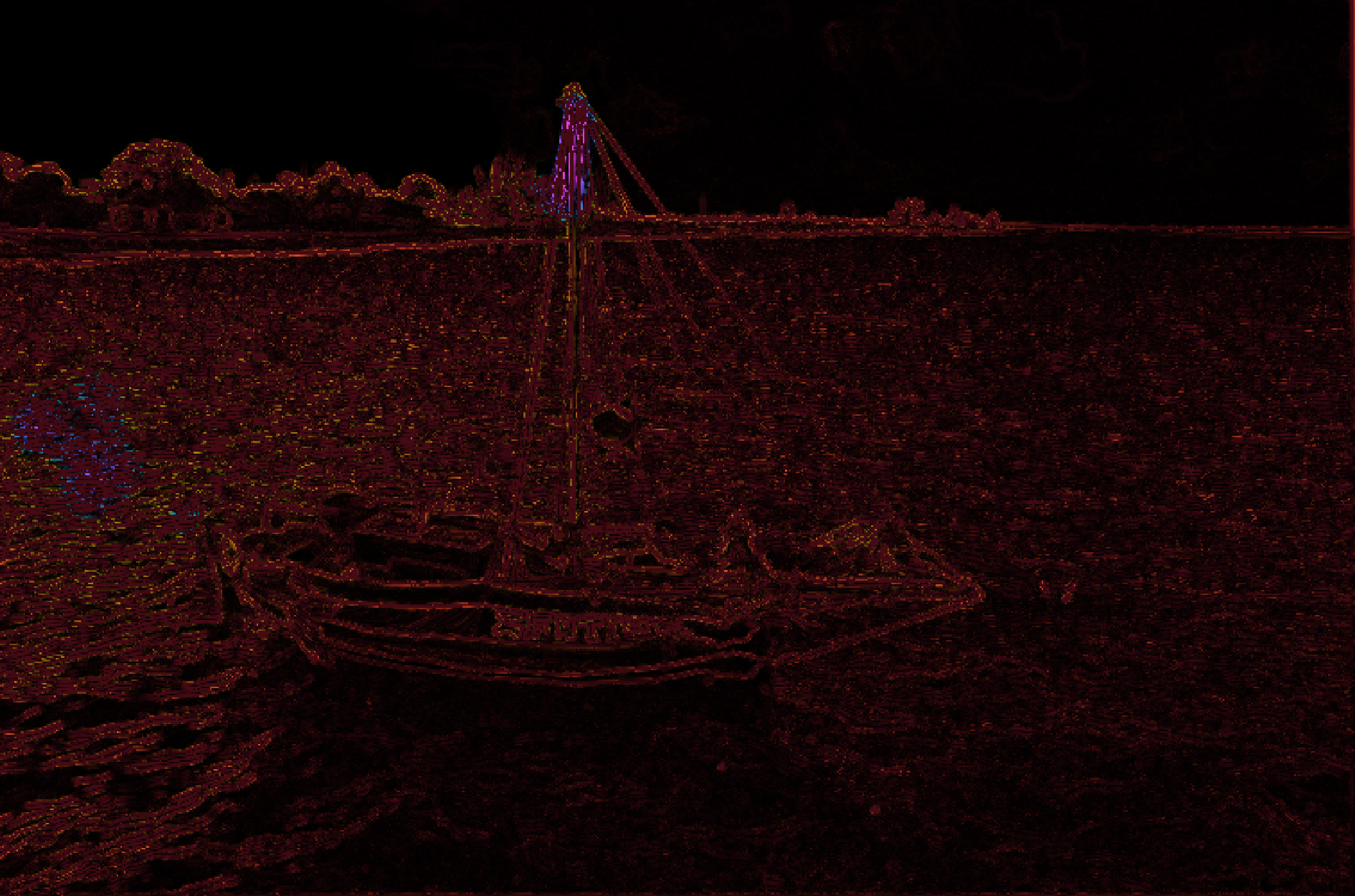}
	\includegraphics[width=1.3in]{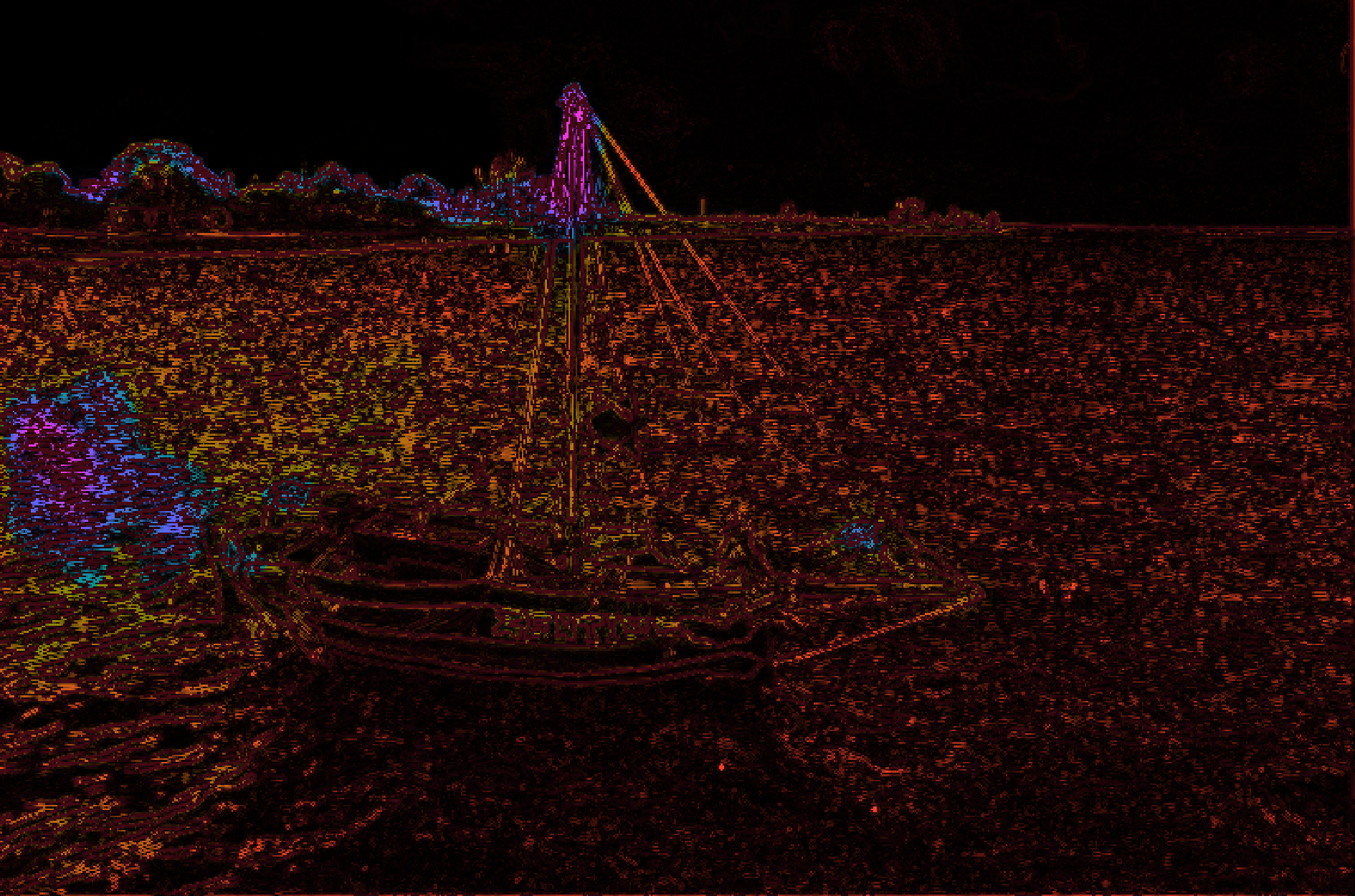}
	\includegraphics[width=1.3in]{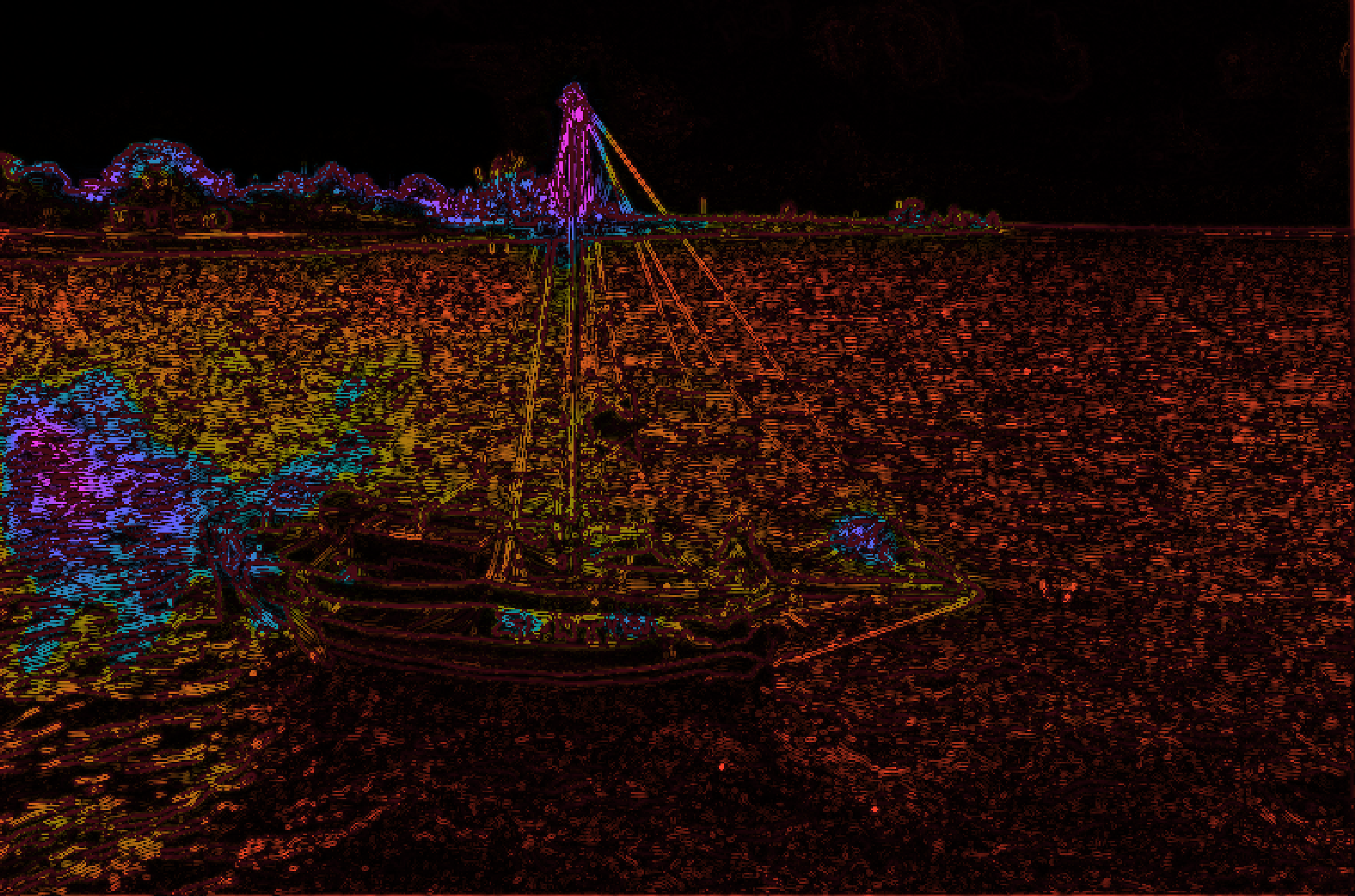}
	\includegraphics[width=1.3in]{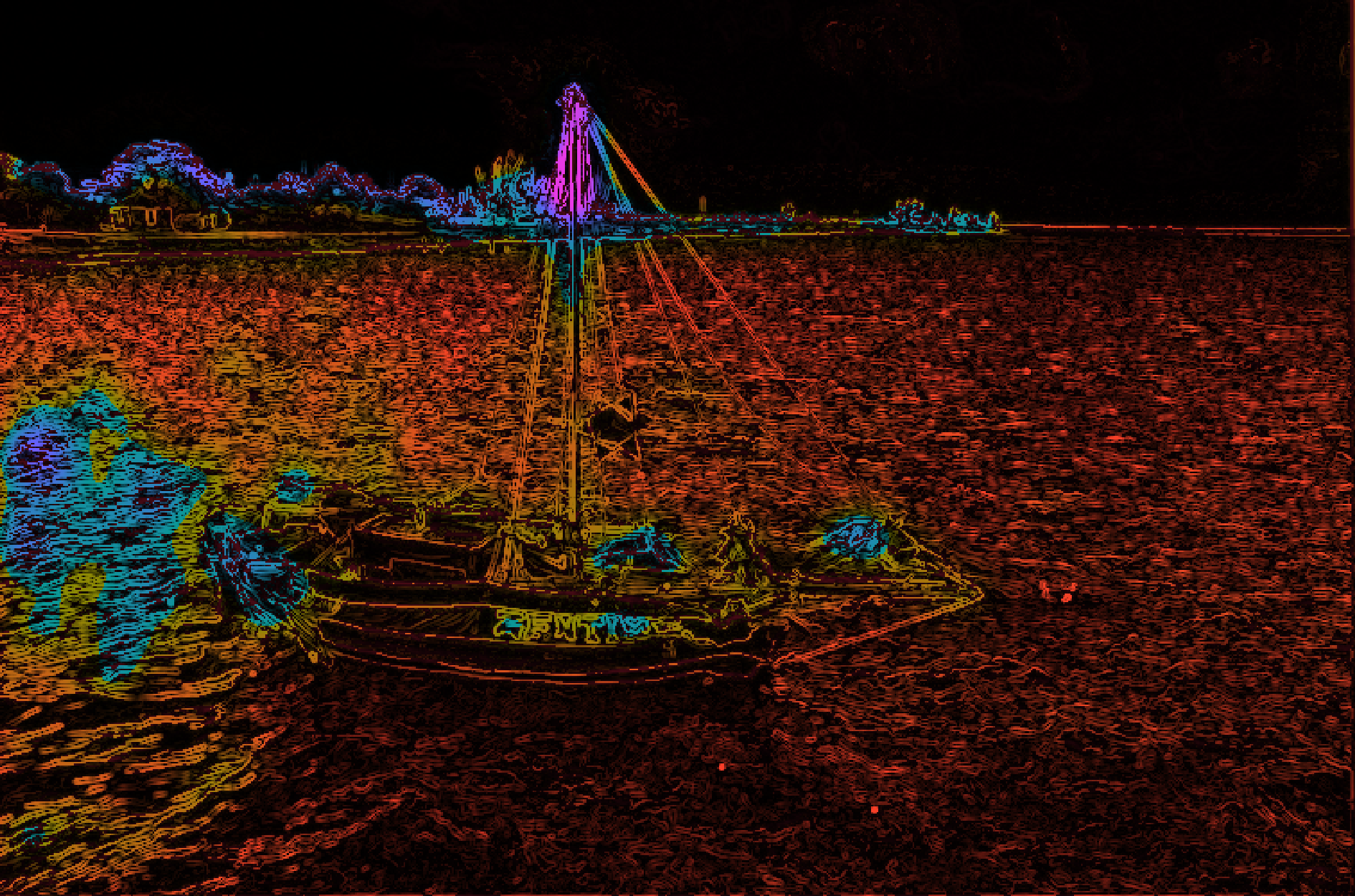}
	\caption{Upper row: Gaussian blurred images "Sailing1" from the LIVE DBR2 dataset, evaluated at Gaussian blur levels of $s_B = 0.5, 0.9, 1.4, 1.9, 3.3$. Lower row: $\lambda_{err}(\mathbf{p}) \Big|_{\Omega_H}$. Blue and violet colors indicate a growing impact of strong edge degradation on texture regions, whereas a more uniform red hue corresponds to mild, barely perceptible losses. Additionally, we can observe where the degradation impact is strongest — specifically in the surrounding regions where strong edges progressively merge as blur increases.}
	\label{fig:sailing1}
\end{figure*}

To estimate the texture degradation in a perceptually meaningful way, the proposed method focuses on measuring local energy loss in the gradient domain, but only within regions of the image that exhibit reliable structural information. This is crucial to avoid spurious responses in textured or noise-sensitive areas.

The first step involves identifying what we define as the \emph{certainty region} — a spatial mask that delimits the structurally stable zones of the image. These are typically formed by strong but isolated edges, where the local image structure is stable and thus suitable for reference-based comparisons. The certainty map $M(\mathbf{p})$ is computed at each pixel location $\mathbf{p}$ as the ratio between the absolute value of the distorted complex gradient field $y(\mathbf{p})$ and the magnitude of the reference complex gradient field $\tilde{y}(\mathbf{p})$:

$$
M(\mathbf{p}) = \frac{|y(\mathbf{p})|}{|\tilde{y}(\mathbf{p})|} \; .
$$

The certainty map defines two mutually exclusive subsets of the gradient domain $\Omega$: the \emph{certainty region} $\Omega_C$ of structurally reliable pixels, and the \emph{weaker-edges region} $\Omega_H$ where weak or overlapping edges dominate:
\begin{IEEEeqnarray}{l}
	\Omega_C = \left\{ \mathbf{p} \in \Omega \ \big| \ M(\mathbf{p}) \geq \overline{M} \right\}, \nonumber \\
	\Omega_H = \left\{ \mathbf{p} \in \Omega \ \big| \ M(\mathbf{p}) < \overline{M} \right\}, \nonumber
\end{IEEEeqnarray}
with threshold $\overline{M} = \sqrt{0.3}$ (a $-10$ dB drop from the peak gradient magnitude), chosen empirically to isolate stable edges while excluding saturated or noisy areas.

Within the region $\Omega_C$, we compute a localized energy difference map $y_{err}(\mathbf{p})$, capturing the structural attenuation between the distorted and reference complex gradients. This is formulated as:
$$
y_{err}(\mathbf{p}) = y(\mathbf{p}) \Big|_{\Omega_C} - \tilde{y}(\mathbf{p}) \Big|_{\Omega_C} \; .
$$

The comparison is strictly localized to $\Omega_C$, where the edge signal is strong and stable, avoiding unreliable responses from weaker or overlapping edges.

The smoothed gradient energy of the error detail is defined as:
\begin{equation}
	\lambda_{err}(\mathbf{p}) = \sum_{\mathbf{q}}{w_\mathbf{p}(\mathbf{q})^2\left|y_{err}\left(\mathbf{p}-\mathbf{q}\right)\right|^2} \; ,
	\label{eqn:lamda_err}
\end{equation}
where $w_\mathbf{p}(\mathbf{q})$ is a sampling window centered on $\mathbf{p}$. In Fig. \ref{fig:sailing1}, we show $\lambda_{err}(\mathbf{p}) \Big|_{\Omega_H}$ for the image "Sailing1" from the LIVE DBR2 dataset, across increasing Gaussian blur levels with $s_B = 0.5, 0.9, 1.4, 1.9, 3.3$. As the blur intensifies, we observe a progressive shift in the residual map from red to blue and violet tones, indicating a growing impact of strong edge degradation on texture regions.

The resulting error map is restricted to the weaker edges region $\Omega_H$, and a single scalar value $\mu$ is derived by averaging the smoothed energy error within this region. Formally, we define:
$$
\mu = \sqrt{\frac{1}{|\Omega_C|} \sum_{\mathbf{p} \in \Omega_H} \lambda_{err}(\mathbf{p})} \; .
$$

Note that the error energy is accumulated over the weaker-edges region $\Omega_H$, where degradations are perceptually most visible, but normalized by the cardinality of $\Omega_C$, not of $\Omega_H$. This asymmetry encodes the assumption that the perceived error injected into the texture is causally tied to the strength of the degraded structural sources in $\Omega_C$, rather than to the extent of $\Omega_H$ over which it is observed.

Unlike typical reduced-reference models that rely on pre-extracted global statistics, this estimation leverages both local filtering and spatial masking, adapted dynamically to each image content. Crucially, $\mu$ is computed once per image pair and serves as a structural prior for the second stage of the system, which operates without further reference access.

A deeper look into the definition of the scalar noise estimate $\mu$ reveals an intentional cross-regional interaction between structurally reliable (cold) and perceptually ambiguous (hot) zones. Specifically, the error energy is measured by evaluating the divergence between the distorted and reference complex gradients only within the structurally reliable set $\Omega_C$ — regions characterized by strong, isolated edges. This ensures that the source of the error is grounded in areas where edge information is unambiguous and consistent across distortions.

However, the impact of this error is not assessed locally: the error energy is propagated through a smoothing kernel $w_\mathbf{p}(\mathbf{q})$ and measured in the surrounding, perceptually more sensitive areas $\Omega_H$ (flat textures, weak edges, overlapping structures), where visual masking is lower. This dual-region approach reflects the intuition that distortions originating in strong edges bleed into neighboring textured zones, amplifying their perceived degradation.

For statistical consistency and invariance to the local geometric support, $\mu$ is normalized by the cardinality of $\Omega_C$ rather than $\Omega_H$, consistently with the perceptual observation that interference from high-contrast structures dominates subjective judgments in nearby textured or flat zones.

This propagation mechanism is conceptually aligned with Marr's theory of early vision, particularly the \emph{raw primal sketch}, where local edge primitives are organized into higher-order representations: the perception of structure depends not only on local gradients but also on their contextual integration, mirrored here by the influence of strong-edge deviations over texture-dominated regions.

Variations in viewing distance are incorporated by adapting the Gaussian kernel scale as $\sigma = \sigma_0 \cdot \tau^2$, with $\tau\overset{\Delta}{=}\delta/\delta_0$ (see Sec. \ref{sec:Background}), so that gradient features capture perceptual effects consistently across distances. When the distance is not available as metadata, $\tau$ is recovered through the same regression on Gaussian-blurred images used for the Full-Reference case \cite{GIANNITRAPANI25}.

Once the structural error has been aggregated into the global scalar value $\mu$, we proceed to compute the proposed \emph{texture fidelity index} $Q_t$, which quantifies the perceptual degradation of non-structured, fine-detail regions.

This estimation relies solely on the distorted image. The smoothed gradient energy of the detail $\lambda(\mathbf{p})$ is:
\begin{equation}
	\lambda(\mathbf{p}) = \sum_{\mathbf{q}}{w_\mathbf{p}(\mathbf{q})^2\left|y\left(\mathbf{p}-\mathbf{q}\right)\right|^2} \; .
	\label{eqn:lamda_dist}
\end{equation}

The average signal energy is then computed as:
$$
\sigma_s^2 = \frac{1}{|\Omega_C|} \sum_{\mathbf{p} \in \Omega_C} \lambda(\mathbf{p}) \; ,
$$
where, for simplicity, we reuse the certainty region $\Omega_C$ as a stable support for signal measurement.

The corresponding noise level is given by $\mu$, previously defined as the energy deviation accumulated over $\Omega_H$ and normalized by $|\Omega_C|$, computed from the difference between gradient maps of the distorted and reference images.

The texture index $Q_t$ is finally computed as a normalized SNR ratio:
$$
Q_t = \frac{\text{SNR}}{\text{SNR}_0} = \frac{\sigma_s^2}{\mu + C} \cdot \frac{C}{\sigma_s^2} = \frac{C}{\mu + C} \; ,
$$
where $C = 20$ sets the noise floor. This expression yields a perceptually normalized fidelity measure bounded in $(0,1)$, with higher values indicating lower relative noise and higher texture fidelity.

This formulation avoids local fitting or patch-based processing, relying on global statistics from the distorted image and on the reference only to estimate the scalar $\mu$ --- which classifies the approach as \emph{Partial-Reference}, though minimal in data dependency.

\subsection{Structural Degradation Index from Hermite-Based Prediction}
\label{subsec:Structural Degradation Index from Hermite-Based Prediction}

\begin{figure*}[!t]
	\centering
	\includegraphics[width=1.3in]{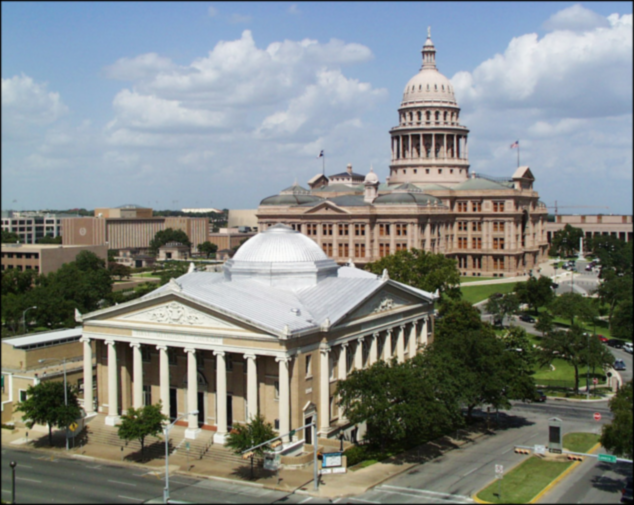}
	\includegraphics[width=1.3in]{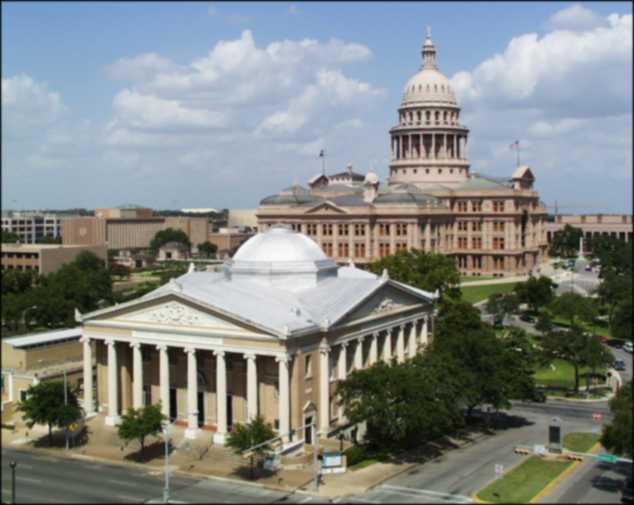}
	\includegraphics[width=1.3in]{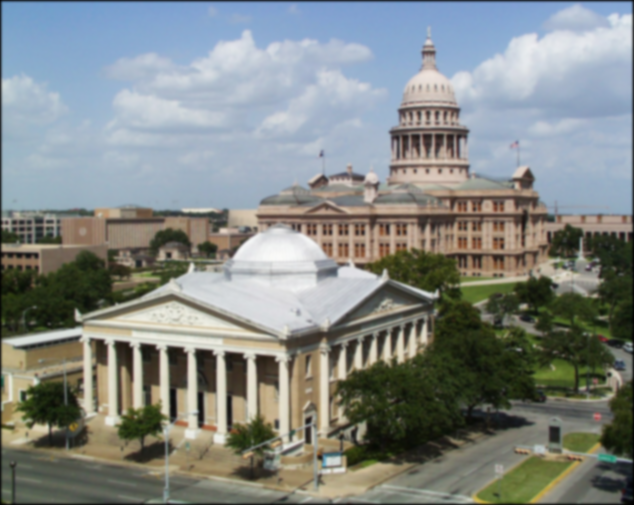}
	\includegraphics[width=1.3in]{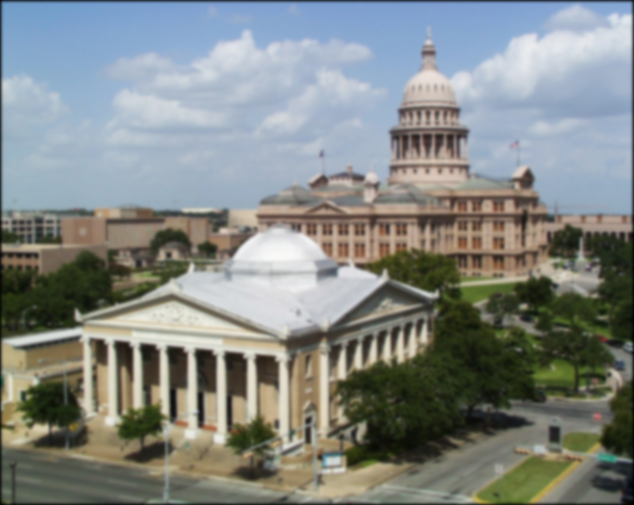}
	\includegraphics[width=1.3in]{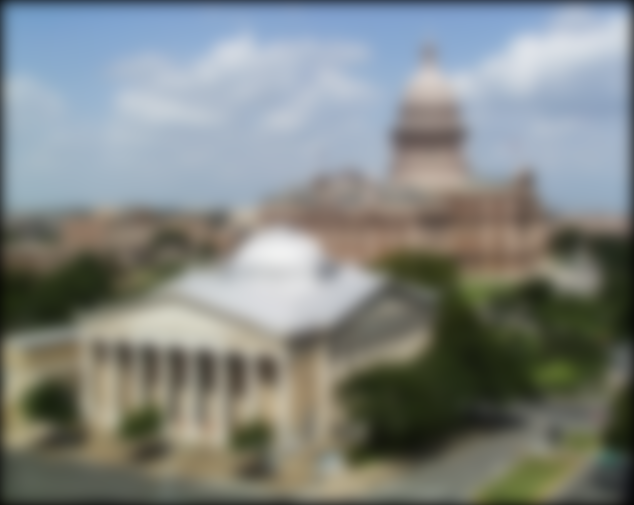} \\
	\vspace{1.5mm}
	\includegraphics[width=1.3in]{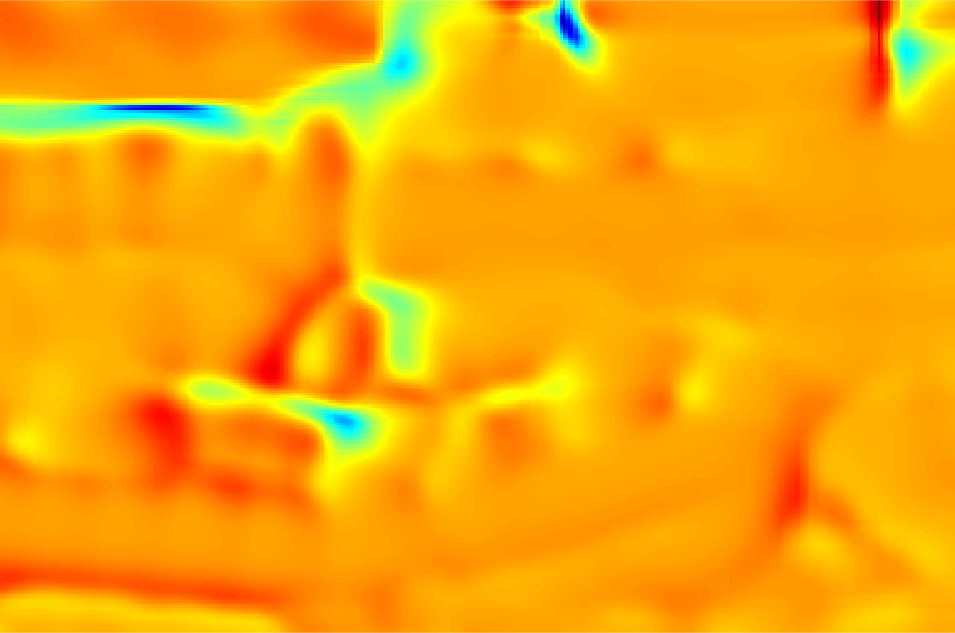}
	\includegraphics[width=1.3in]{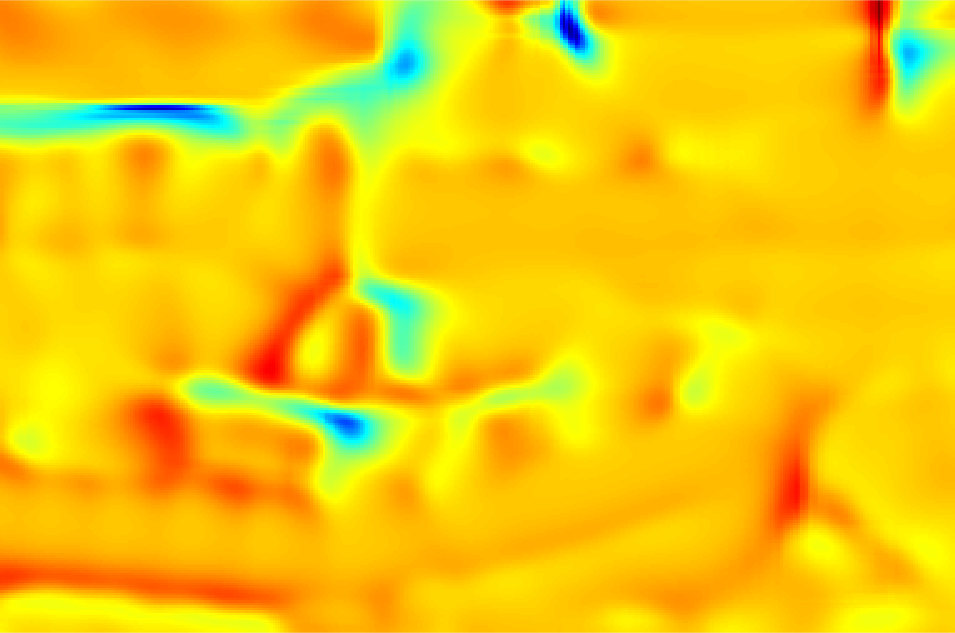}
	\includegraphics[width=1.3in]{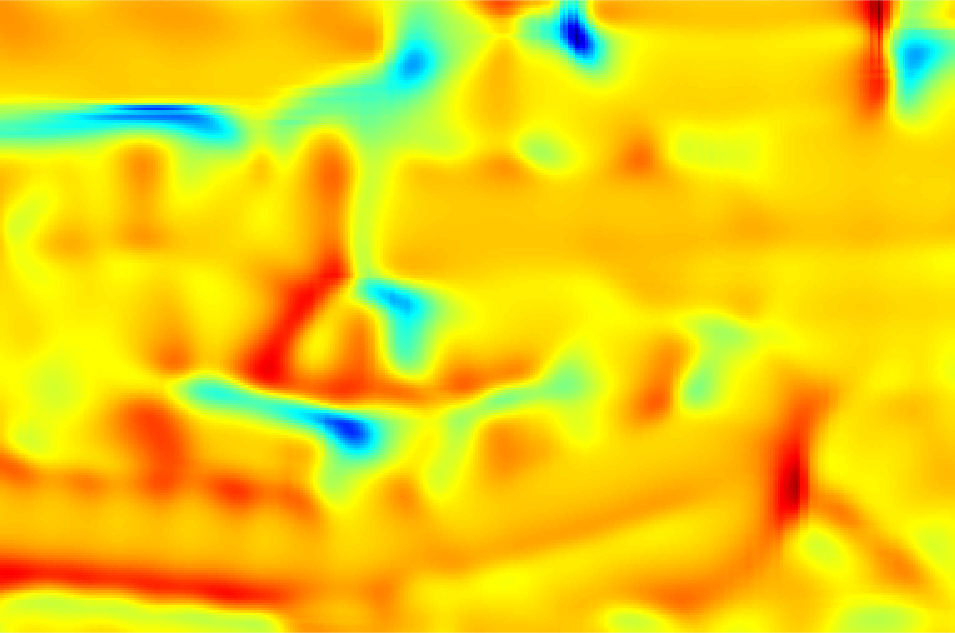}
	\includegraphics[width=1.3in]{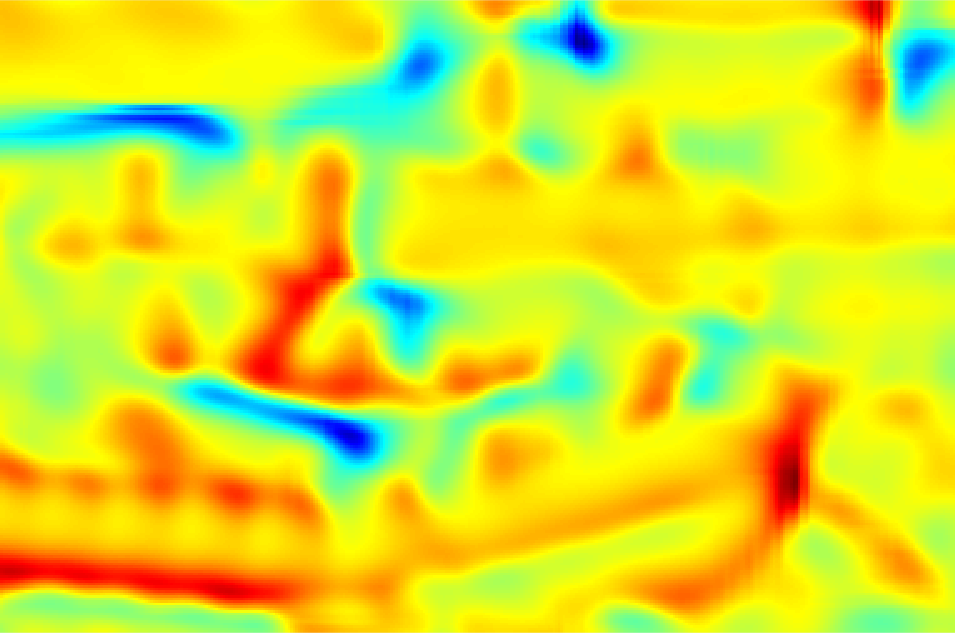}
	\includegraphics[width=1.3in]{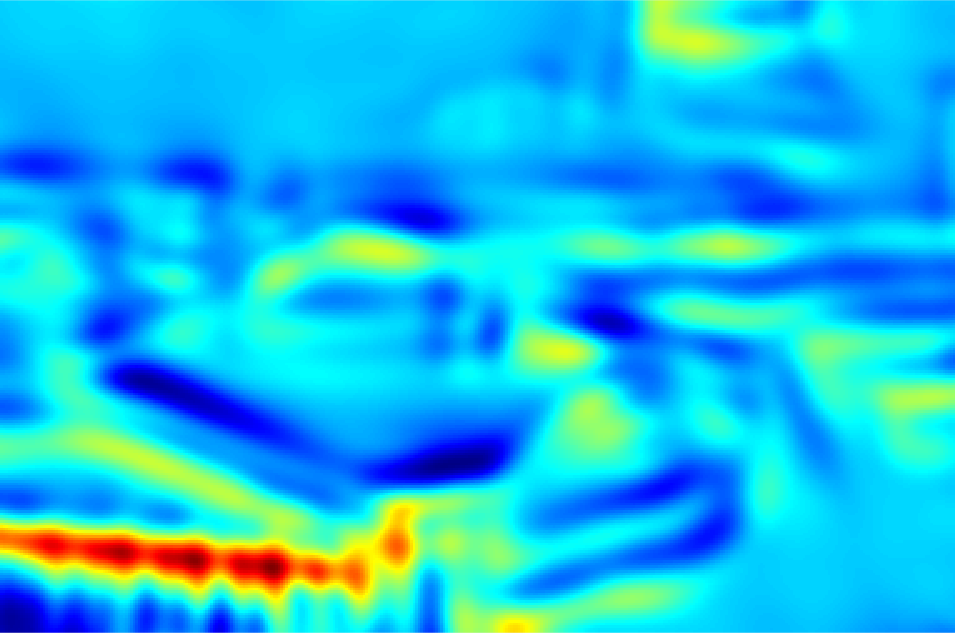}
	\caption{Upper row: Gaussian blurred images "Churchandcapitol" from the LIVE DBR2 dataset, evaluated at Gaussian blur levels of $s_B = 1.0, 1.3, 1.9, 2.5, 10.8$. Lower row: $\nabla \cdot \mathbf{G(\mathbf{p})}$. As blur increases, closely spaced strong edges progressively merge, causing a more pronounced impact on perceived image degradation compared to other regions, which remain mostly light blue. The wide Gaussian window ($\sigma_w=10$) used for regression emphasizes structural coherence over large areas, highlighting zones of residual curvature where strong edges were previously concentrated as blur increases. Focusing on the central image region — where the visual system is most sensitive — enhances discriminability, consistent with the perceptual dominance of foveal vision and supporting central cropping as a refinement strategy.}
	\label{fig:churchandcapitol}
\end{figure*}

The structural analysis module is founded on the use of Hermite-Gauss filters to extract a local geometric description of the distorted gradient map in the absence of an explicit reference. This is accomplished through a self-prediction mechanism that allows estimation of structural regularity and coherence by leveraging internal correlations within the distorted content.

Let $y(\mathbf{p})$ be the \emph{complex gradient map} of the distorted image, from which we extract local patches centered at pixel location $\mathbf{p}$. These patches are projected onto a set of second-order Hermite-Gauss basis functions, specifically $\phi_{20}$ and $\phi_{02}$, designed to capture directional curvature in the horizontal and vertical axes:
\begin{equation}
	\phi_2\left(x_i\right) \equiv \left[\sqrt2 \frac{x_i^2}{\sigma^2}-\frac{1}{\sqrt2}\right]\pi^{-1/4}\sigma^{-1}e^{-\textstyle\frac{x_i^2}{2\sigma^2}} \; ,
\end{equation}
where $x_i=x_1$ and $x_i=x_2$ are the spatial coordinates of the map and $\sigma$ is the standard deviation of the Gaussian kernel modified in distance as in \ref{subsec:Partial-Reference Texture Index}.

Each patch is thus encoded as a weighted combination of these basis components using a weighted least-squares regression:
$$
B(\mathbf{p}) = \left(R^\top W R\right)^{-1} R^\top W X(\mathbf{p}) \; ,
$$
where $R$ contains the vectorized Hermite-Gauss basis sampled over the local window, $X(\mathbf{p})$ is the observed gradient patch, and $W$ is a Gaussian weighting matrix enforcing locality. The coefficients $B_2(\mathbf{p})$ and $B_3(\mathbf{p})$, i.e. the local strengths of the $(2,0)$ and $(0,2)$ filters, form the surrogate structural field:
$$
\mathbf{G}(\mathbf{p}) = 
\begin{bmatrix}
	B_2(\mathbf{p}) \\
	B_3(\mathbf{p})
\end{bmatrix}.
$$

Unlike classical gradients, this field is derived entirely from the distorted image, making the analysis fully \emph{No-Reference}. The key insight is that $\mathbf{G}(\mathbf{p})$ encapsulates local curvature and flow orientation, enabling a perceptual characterization of structural integrity.

To quantify structural distortion, we compute the divergence of this field over the entire image:
$$
\nabla \cdot \mathbf{G(\mathbf{p})} = \frac{\partial B_2(\mathbf{p})}{\partial x_1} + \frac{\partial B_3(\mathbf{p})}{\partial x_2} \; ,
$$
which acts as a \emph{vector Laplacian} of the complex gradient field $y(\mathbf{p})$, measuring local expansion or compression in the directional flow and thus highlighting geometric inconsistencies. As illustrated in Fig. \ref{fig:churchandcapitol} (image ``Churchandcapitol'', LIVE DBR2, at increasing blur levels), the Laplacian magnitude rises where closely spaced strong edges merge, providing a sensitive indicator of structural degeneration that matches the perceived quality decline.

In parallel, we evaluate the \emph{angular variability} of the structural field by computing the variance of the orientation angle:
$$
\theta(\mathbf{p}) = \operatorname{atan2}(B_3(\mathbf{p}), B_2(\mathbf{p})) \; , \quad
\text{Var}_\theta = \text{Var}(\theta(\mathbf{p})) \; .
$$

The \emph{total curvature measure} is then defined as a weighted combination:
$$
\mathcal{E}_{\text{total}}(\mathbf{p}) = \left\langle |\nabla \cdot \mathbf{G}| \right\rangle_{\mathbf{p}} + \gamma_0 \cdot \text{Var}_\theta \; ,
$$
where $\left\langle \cdot \right\rangle_{\mathbf{p}}$ denotes averaging over the pixels of the local patch indexed by $\mathbf{p}$, over which $\text{Var}_\theta$ is likewise computed; the divergence term captures the breakdown of local structure, $\text{Var}_\theta$ the orientation incoherence, and $\gamma_0 = 10$ balances the two.

The resulting \emph{structural reliability weight} is obtained through a single exponential of unit rate,
$$
w(\mathbf{p}) = \exp\!\left(-\gamma_1 \cdot \mathcal{E}_{\text{total}}(\mathbf{p})\right) \; , \quad \gamma_1 = 1 \; ,
$$
mapping the local irregularity into a reliability factor in $(0,1]$: coherent regions ($\mathcal{E}_{\text{total}} \!\to\! 0$) approach $w \!\to\! 1$, ambiguous ones are suppressed ($w \!\to\! 0$). The unit rate $\gamma_1 = 1$ is the natural, unscaled modulation, leaving $\gamma_0$ as the only free constant; both are fixed across all datasets, with a sensitivity analysis in the supplementary material confirming robustness to $\gamma_0$.

We then adopt an \emph{information-theoretic interpretation}: quality evaluation is framed as a transmission problem over a \emph{self-channel}, where the distorted image acts as both input and output and degradation is modeled as additive noise. Unlike VIF, no reference is available, so both source and observation derive from the same distorted observation $\tilde{x}(\mathbf{p}) = x(\mathbf{p}) = |y(\mathbf{p})|$. The channel signal-to-noise relationship is modulated by two quantities: $\mu$, the \emph{residual noise estimate} derived from structural deviations in weakly textured regions ($\Omega_H$); and $\mathcal{E}_{\text{total}}$, which governs both the reliability weight $w$ and the noise amplification.

The \emph{modulated noise variance} is obtained by amplifying the base noise level $\mu^2$ through the inverse of the structural reliability,
$$
\sigma_v^2(\mathbf{p}) = \frac{\mu^2}{w(\mathbf{p})} = \mu^2 \cdot \exp\!\left(\gamma_1 \cdot \mathcal{E}_{\text{total}}(\mathbf{p})\right) \; ,
$$
tying the two roles of $\mathcal{E}_{\text{total}}$ to a single rate $\gamma_1$: reliable regions ($w \!\to\! 1$) yield $\sigma_v^2 \!\to\! \mu^2$, while severely damaged ones ($w \!\to\! 0$) diverge ($\sigma_v^2 \!\to\! \infty$), driving the retained information to zero; the exponential form also guarantees $\sigma_v^2 > 0$. The final score is a perceptually weighted, log-domain information ratio:
$$
Q_{\text{struct}} = \frac{
	\sum_{\mathbf{p}} w(\mathbf{p}) \, \log_2 \left(1 + \frac{g^2 \cdot \sigma_x^2}{\sigma_v^2 + \sigma_n^2} \right)
}{
	\sum_{\mathbf{p}} w(\mathbf{p}) \, \log_2 \left(1 + \frac{\sigma_x^2}{\sigma_n^2} \right)
} \; ,
$$
where $\sigma_x^2$ is the local signal energy derived from $|y(\mathbf{p})|$, $\sigma_n^2$ is a constant noise floor, and $g = \frac{\text{Cov}(\tilde{x}, x)}{\text{Var}(\tilde{x}) + \varepsilon} \approx 1$ reflects unit-gain transmission. The same reliability weight $w(\mathbf{p})$ multiplies both numerator and denominator, acting as a perceptual aggregation weight that emphasizes structurally coherent regions, in the spirit of the sub-band weighting of the original VIF. The weighted sums run over all image patches, yielding a score that reflects how much structural information is preserved under implicit self-degradation.

The structural quality term $Q_{\text{struct}}$ can be interpreted as a \emph{normalized information capacity} between predicted and original image structures: a ratio of two log-domain signal-to-noise terms, akin to a channel capacity under Gaussian noise, quantifying the structural information retained relative to an undistorted scenario.

This formulation is fully blind, relying solely on internal consistency within the distorted image, yet perceptually grounded through both signal statistics and structural geometry. The scalar $\mu$ adapts the model to the noise context of each image, enabling stable prediction across distortion types.

The divergence map $\nabla \cdot \mathbf{G(\mathbf{p})}$, i.e. the local Laplacian of the complex gradient field $y(\mathbf{p})$, derived from smooth structural coefficient maps $B(\mathbf{p})$, offers a robust structural regularity descriptor, even in the absence of a reference image. The wide Gaussian window $\sigma_{w}=10$ used for regression enforces structural integration over extended spatial supports, making the map responsive to regional consistency rather than local detail.

Focusing the analysis on the central image region, where the visual system is most sensitive to structural distortions, further improves the discriminative power of the index, consistently with the dominance of foveal vision in perceptual evaluation.

\subsection{Final Quality Prediction}
\label{subsec:Final Quality Prediction}

\noindent The two indices derived in the previous subsections capture complementary aspects of perceived degradation: the texture index $Q_t$ measures the perceptual loss in fine, non-structured detail, while the structural index $Q_{\text{struct}}$ quantifies the integrity of the dominant geometric structures. Crucially, both indices are computed on operators whose scale is set by the same viewing-distance-dependent kernel $\sigma = \sigma_0 \cdot \tau^2$, so that they are already expressed on a common perceptual scale. This allows the two scores to be merged through a simple low-order combination, without reintroducing the viewing distance at the fusion stage and without the VQEG five-parameter logistic rectification \cite{VQEG00}.

The final quality score is obtained as a first-order (affine) combination of the two indices:
\begin{equation}
	\text{PreSPA} = b_0 + b_1 \cdot a \cdot Q_{\text{struct}} + b_2 \cdot a \cdot Q_t \; ,
	\label{eqn:prespa}
\end{equation}
which depends on three model parameters --- the coefficients $(b_0, b_1, b_2)$ controlling the affine fusion --- together with a pair $(a, \tau)$ that is fixed a priori from metadata or blur regression and, following the same convention as in BELE \cite{GIANNITRAPANI25}, is not included in the parameter count. No parameter is learned through backpropagation or stochastic optimization.

The two a priori parameters $(a, \tau)$ mirror the role of the pair $(Q, \tau)$ in the Full-Reference estimator BELE \cite{GIANNITRAPANI25, giannitrapani2025beleblurequivalentlinearized} and are not fitted on subjective scores, being retrieved from metadata or estimated through regression on Gaussian-blurred images. Here $a$ is a statistical anchor that scales the scoring system to the database, setting the lower bound of perceived quality \cite{DICLAUDIO21}, while $\tau \overset{\Delta}{=} \delta/\delta_0$ is the normalized viewing distance, incorporated into the scale of both indices as $\sigma = \sigma_0 \cdot \tau^2$ (Sec.~\ref{sec:Background}).

The three coefficients $(b_0, b_1, b_2)$ arise from the fusion in Eq.~(\ref{eqn:prespa}): $b_0$ is a bias term, while $b_1$ and $b_2$ weight the structural and textural components. They are obtained, for each dataset, through the same robust weighted affine regression against the DMOS scale adopted in \cite{GIANNITRAPANI25}; since the combination is affine and the indices are already perceptually aligned, this fitting is a per-dataset scale-and-offset alignment, not a data-driven training stage.

Together with the reliance on spatial-domain operations only, this keeps PreSPA interpretable, generalizable, and computationally lightweight, as shown in the next section.

\section{Experiments}
\label{sec:Experiments}

\noindent The PreSPA framework requires no training and no dataset-specific hyperparameter tuning: the three model parameters $(b_0, b_1, b_2)$ perform the affine alignment to the DMOS scale of each dataset in place of the VQEG logistic rectification, while the two constants $(a, \tau)$ are fixed a priori as in BELE \cite{GIANNITRAPANI25} and not counted among the model parameters, as detailed in Sec.~\ref{subsec:Final Quality Prediction}. This makes PreSPA particularly suited for evaluation across heterogeneous benchmarks, since the same model is applied unchanged to all datasets. In this section we report a comprehensive experimental campaign on six widely adopted IQA benchmarks. We first present per-distortion results against state-of-the-art No-Reference (NR) methods (Sec.~\ref{subsec:Performance Evaluation}), then summarise the overall performance and contextualise PreSPA against the full landscape including Full-Reference (FR) methods (Sec.~\ref{subsec:Overall Performance and FR Comparison}), and finally analyse the computational cost (Sec.~\ref{subsec:Computational Efficiency}).

\subsection{Datasets and Protocols}
\label{subsec:Datasets and Protocols}

\noindent The evaluation spans six publicly available datasets covering a wide spectrum of distortions: \emph{LIVE DBR2} \cite{SHEIKH06B} ($779$ images, $5$ classic distortions), \emph{TID2013} \cite{PONOMARENKO15} ($3000$ images, $24$ types including chromatic and contrast/saturation changes), \emph{CSIQ} \cite{LARSON10} ($866$ images, six classes), \emph{LIVE MD} \cite{JAYARAMAN12} ($450$ multiply-distorted images), \emph{KADID-10K} \cite{LIN20} ($10125$ images, $25$ types including colour-domain artefacts), and \emph{PIPAL} \cite{PIPAL20} ($21800$ images, restoration- and GAN-induced distortions). For all datasets, $\tau$ and $a$ are retrieved from metadata or recovered through the regression procedure of \cite{GIANNITRAPANI25}, while the three fusion coefficients $(b_0, b_1, b_2)$ are obtained on each dataset through the same robust weighted affine regression used in \cite{GIANNITRAPANI25}, an alignment strictly less flexible than the five-parameter VQEG logistic.

PreSPA is compared against deep NR methods (TOPIQ-NR \cite{CHEN24}, MUSIQ \cite{KE21}, HyperIQA \cite{SU20}, CNNIQA \cite{KANG14CNNIQA}, CLIPIQA and CLIPIQA+ \cite{WANG23}), traditional NSS-based NR methods (BRISQUE \cite{MITTAL12}, NIQE \cite{MITTAL13B}), deep FR methods (TOPIQ-FR \cite{CHEN24}, TOPIQ-FR-PIPAL \cite{PIPAL20}, DISTS \cite{DING22}, LPIPS and LPIPS-VGG \cite{ZHANG18}, PIEAPP \cite{PRASHNANI18}), and classical FR metrics (VIF \cite{SHEIKH06}, MS-SSIM \cite{WANG03}, FSIM \cite{ZHANG11}, GMSD \cite{XUE14}). Performance is measured by SROCC and PLCC against subjective DMOS; all NR competitors are re-aligned on each dataset's DMOS scale prior to PLCC computation to ensure a fair comparison.

\subsection{Performance Evaluation Across Distortion Types}
\label{subsec:Performance Evaluation}

\begin{table*}[!t]
	\scalebox{1}{
		\begin{minipage}{\textwidth}
			\captionsetup{width=\textwidth}
			\caption{Experimental verification of PreSPA compared to NR Deep and Clip-based methods on the full LIVE DBR2, TID2013, CSIQ, LIVE MD, KADID-10K, and PIPAL datasets. Best values for each distortion are highlighted in bold blue, and second-best values in bold black. The table also includes computational cost (FLOPS) and the number of parameters used by each method. FLOPS refers to the computational cost per pair of processed images, with deep metric values expressed in gigaflops (GFLOPS). The number of parameters for deep metrics is expressed in millions (M).}
			\label{tab:classicIQAdatasetsComparison}
			
			\renewcommand{\arraystretch}{1.0} 
			\begin{adjustbox}{max width=1.0\textwidth}
				\begin{tabularx}{1.36\textwidth}{llcccccccccccccc}
					
					\toprule
					\multicolumn{2}{c}{} & \multicolumn{2}{c}{TOPIQ-NR} & \multicolumn{2}{c}{HyperIQA} & \multicolumn{2}{c}{MUSIQ} & \multicolumn{2}{c}{CNNIQA} & \multicolumn{2}{c}{CLIPIQA} & \multicolumn{2}{c}{CLIPIQA+} & \multicolumn{2}{c}{PreSPA} \\
					\midrule
					& Parameter no. & \multicolumn{2}{c}{$\scriptstyle 45$ M} & \multicolumn{2}{c}{$\scriptstyle 27$ M} & \multicolumn{2}{c}{$\scriptstyle 27$ M} & \multicolumn{2}{c}{$\scriptstyle 0.7$ M} & \multicolumn{2}{c}{$\scriptstyle 102$ M} & \multicolumn{2}{c}{$\scriptstyle 102$ M} & \multicolumn{2}{c}{$\scriptstyle 3$} \\
					& GFLOPS & \multicolumn{2}{c}{$\scriptstyle 14$} & \multicolumn{2}{c}{$\scriptstyle 215$} & \multicolumn{2}{c}{$\scriptstyle 17$} & \multicolumn{2}{c}{$\scriptstyle 0.7$} & \multicolumn{2}{c}{$\scriptstyle 35$} & \multicolumn{2}{c}{$\scriptstyle 12$} & \multicolumn{2}{c}{$\scriptstyle 0.13$} \\
					
					\midrule
					Traditional & \multirow{2}{*}{Distortion type} & \multirow{2}{*}{SROCC} & \multirow{2}{*}{PLCC} & \multirow{2}{*}{SROCC} & \multirow{2}{*}{PLCC} & \multirow{2}{*}{SROCC} & \multirow{2}{*}{PLCC} & \multirow{2}{*}{SROCC} & \multirow{2}{*}{PLCC} & \multirow{2}{*}{SROCC} & \multirow{2}{*}{PLCC} & \multirow{2}{*}{SROCC} & \multirow{2}{*}{PLCC} & \multirow{2}{*}{SROCC} & \multirow{2}{*}{PLCC} \\
					datasets &  & \multicolumn{2}{c}{} & \multicolumn{2}{c}{} & \multicolumn{2}{c}{} & \multicolumn{2}{c}{} & \multicolumn{2}{c}{} & \multicolumn{2}{c}{} & \multicolumn{2}{c}{} \\
					\midrule
					\multirow{5}{*}{LIVE DBR2} & Gaussian Blur & 0.93274 & \textbf{0.89725} & 0.86608 & 0.82924 & \textbf{0.94053} & 0.87463 & 0.76188 & 0.39043 & 0.76852 & 0.72046 & 0.78398 & 0.82801 & $\textcolor{blue}{\textbf{0.96048}}$ & $\textcolor{blue}{\textbf{0.95504}}$ \\
					& Bit Errors in JPEG2000 Stream & 0.92124 & \textbf{0.91702} & 0.91458 & 0.91450 & \textbf{0.92287} & 0.91687 & 0.62370 & 0.49036 & 0.78977 & 0.79767 & 0.90138 & 0.89911 & $\textcolor{blue}{\textbf{0.94607}}$ & $\textcolor{blue}{\textbf{0.94543}}$ \\
					& JPEG Compression & 0.89448 & 0.89514 & 0.86272 & 0.86872 & \textbf{0.94071} & \textbf{0.94239} & 0.21366 & 0.24275 & 0.85137 & 0.87511 & 0.89707 & 0.93263 & $\textcolor{blue}{\textbf{0.98533}}$ & $\textcolor{blue}{\textbf{0.98527}}$ \\
					& JPEG2000 Compression & \textbf{0.94142} & \textbf{0.94430} & 0.93299 & 0.93901 & 0.92500 & 0.92521 & 0.54915 & 0.52538 & 0.74366 & 0.73839 & 0.85729 & 0.85674 & $\textcolor{blue}{\textbf{0.97151}}$ & $\textcolor{blue}{\textbf{0.96791}}$ \\
					& Gaussian White Noise & 0.70972 & 0.69436 & \textbf{0.75871} & \textbf{0.76188} & 0.69429 & 0.67507 & 0.08750 & 0.09156 & 0.41509 & 0.42200 & 0.71287 & 0.69317 & $\textcolor{blue}{\textbf{0.98666}}$ & $\textcolor{blue}{\textbf{0.98444}}$ \\
					\midrule
					\multirow{24}{*}{TID2013} & Colour Additive Noise & 0.62682 & 0.58934 & 0.62323 & 0.52779 & \textbf{0.63579} & \textbf{0.62581} & 0.02180 & 0.10084 & 0.58894 & 0.54843 & 0.57387 & 0.59658 & $\textcolor{blue}{\textbf{0.83960}}$ & $\textcolor{blue}{\textbf{0.86136}}$ \\
					& Gaussian Blur & \textbf{0.92505} & \textbf{0.92144} & 0.82048 & 0.77698 & 0.89066 & 0.90100 & 0.69940 & 0.66754 & 0.85411 & 0.85297 & 0.77795 & 0.77284 & $\textcolor{blue}{\textbf{0.95309}}$ & $\textcolor{blue}{\textbf{0.95117}}$ \\
					& Gaussian White Noise & 0.65971 & 0.62273 & 0.71239 & 0.64655 & \textbf{0.77340} & \textbf{0.75927} & 0.04828 & 0.10156 & 0.70355 & 0.68855 & 0.72446 & 0.72552 & $\textcolor{blue}{\textbf{0.90840}}$ & $\textcolor{blue}{\textbf{0.91274}}$ \\
					& High Frequency Noise & 0.74507 & 0.75907 & 0.75576 & 0.74185 & \textbf{0.84673} & \textbf{0.86209} & 0.26863 & 0.33859 & 0.54571 & 0.50488 & 0.68474 & 0.72176 & $\textcolor{blue}{\textbf{0.89572}}$ & $\textcolor{blue}{\textbf{0.94581}}$ \\
					& Impulse Noise & 0.64904 & 0.65855 & 0.66041 & 0.59586 & \textbf{0.77825} & \textbf{0.74452} & 0.27831 & 0.32797 & 0.36169 & 0.32149 & 0.47923 & 0.46921 & $\textcolor{blue}{\textbf{0.89086}}$ & $\textcolor{blue}{\textbf{0.88519}}$ \\
					& Masked Noise & \textbf{0.78315} & \textbf{0.79147} & 0.66655 & 0.55274 & 0.60153 & 0.61748 & 0.34075 & 0.34582 & 0.44148 & 0.42385 & 0.50980 & 0.56311 & $\textcolor{blue}{\textbf{0.85233}}$ & $\textcolor{blue}{\textbf{0.87252}}$ \\
					& Quantization Noise & 0.25480 & 0.23119 & 0.40000 & 0.31475 & 0.47582 & 0.48990 & 0.06384 & 0.02268 & 0.54508 & 0.53467 & \textbf{0.76459} & \textbf{0.76173} & $\textcolor{blue}{\textbf{0.87260}}$ & $\textcolor{blue}{\textbf{0.86688}}$ \\
					& Spatially Correlated Noise & 0.71235 & 0.69788 & 0.68339 & 0.66422 & 0.74179 & 0.72837 & 0.11773 & 0.07067 & 0.78217 & 0.77537 & \textbf{0.79058} & \textbf{0.78298} & $\textcolor{blue}{\textbf{0.90806}}$ & $\textcolor{blue}{\textbf{0.90997}}$ \\
					& Block-wise Distortions & 0.10861 & 0.04809 & 0.01898 & 0.00566 & 0.11489 & 0.07802 & \textbf{0.33630} & \textbf{0.22546} & \textcolor{blue}{\textbf{0.46452}} & \textcolor{blue}{\textbf{0.36330}} & 0.20482 & 0.16704 & $0.22538$ & $0.14323$ \\
					& Chromatic Aberrations & \textbf{0.85052} & 0.86300 & 0.84608 & 0.86926 & 0.84705 & \textbf{0.94880} & 0.78702 & 0.80183 & 0.78449 & 0.80556 & 0.76505 & 0.83679 & $\textcolor{blue}{\textbf{0.88302}}$ & $\textcolor{blue}{\textbf{0.95281}}$ \\
					& Comfort Noise & 0.52637 & 0.46556 & 0.65425 & 0.50466 & 0.59043 & 0.59116 & 0.27296 & 0.21507 & 0.64643 & 0.58410 & \textbf{0.66224} & \textbf{0.69523} & $\textcolor{blue}{\textbf{0.91771}}$ & $\textcolor{blue}{\textbf{0.91280}}$ \\
					& Contrast Change & 0.38315 & 0.30702 & 0.15092 & 0.10575 & 0.11899 & 0.10916 & \textcolor{blue}{\textbf{0.62456}} & \textcolor{blue}{\textbf{0.64841}} & \textbf{0.56644} & \textbf{0.49017} & 0.26046 & 0.27556 & $0.42381$ & $0.40812$ \\
					& Image Denoising & 0.86502 & 0.88509 & 0.87680 & 0.88293 & 0.88980 & 0.89933 & 0.76247 & 0.81354 & 0.88704 & 0.87712 & \textbf{0.90000} & \textbf{0.90758} & $\textcolor{blue}{\textbf{0.94791}}$ & $\textcolor{blue}{\textbf{0.96647}}$ \\
					& Dither Color Quantization & 0.56406 & 0.55550 & 0.42716 & 0.35275 & 0.65841 & \textbf{0.67491} & 0.17040 & 0.17670 & 0.11141 & 0.05613 & \textbf{0.68562} & 0.66958 & $\textcolor{blue}{\textbf{0.83708}}$ & $\textcolor{blue}{\textbf{0.84979}}$ \\
					& JPEG Compression & 0.83288 & 0.83450 & 0.83058 & 0.83370 & 0.82563 & 0.86966 & 0.22934 & 0.21217 & \textbf{0.85239} & 0.90430 & 0.84947 & \textbf{0.92415} & $\textcolor{blue}{\textbf{0.96096}}$ & $\textcolor{blue}{\textbf{0.97460}}$ \\
					& JPEG Transmission Errors & 0.58014 & 0.61977 & 0.65794 & 0.56877 & 0.64383 & 0.64622 & 0.30765 & 0.29837 & 0.68109 & 0.74426 & \textbf{0.72652} & \textbf{0.77839} & $\textcolor{blue}{\textbf{0.82290}}$ & $\textcolor{blue}{\textbf{0.81436}}$ \\
					& JPEG2000 Compression & 0.92007 & 0.92863 & 0.91439 & 0.90607 & \textbf{0.92807} & \textbf{0.95152} & 0.82646 & 0.87659 & 0.74987 & 0.80643 & 0.82437 & 0.86985 & $\textcolor{blue}{\textbf{0.96581}}$ & $\textcolor{blue}{\textbf{0.96600}}$ \\
					& JPEG2000 Transmission Errors & 0.64596 & 0.61761 & 0.65726 & 0.51765 & \textbf{0.73904} & \textbf{0.67027} & 0.22183 & 0.09995 & 0.66558 & 0.63552 & 0.59737 & 0.58227 & $\textcolor{blue}{\textbf{0.93760}}$ & $\textcolor{blue}{\textbf{0.92716}}$ \\
					& Lossy Compression & 0.85226 & 0.85179 & 0.83334 & 0.83437 & \textbf{0.89179} & \textbf{0.89514} & 0.46907 & 0.52038 & 0.84867 & 0.87828 & 0.87159 & 0.89070 & $\textcolor{blue}{\textbf{0.95934}}$ & $\textcolor{blue}{\textbf{0.96535}}$ \\
					& Mean Shift & 0.15512 & 0.05539 & 0.13134 & 0.05233 & \textbf{0.27508} & \textbf{0.30010} & 0.17662 & 0.06553 & 0.02427 & 0.01849 & 0.01246 & 0.01676 & $\textcolor{blue}{\textbf{0.73179}}$ & $\textcolor{blue}{\textbf{0.76657}}$ \\
					& Multiplicative Gaussian Noise & 0.57115 & 0.55416 & 0.61375 & 0.49003 & \textbf{0.66181} & \textbf{0.65699} & 0.13238 & 0.16690 & 0.61004 & 0.59071 & 0.64883 & 0.65364 & $\textcolor{blue}{\textbf{0.86141}}$ & $\textcolor{blue}{\textbf{0.86477}}$ \\
					& Non-Eccentricity Pattern Noise & 0.04041 & 0.02820 & 0.03429 & 0.02692 & 0.01720 & 0.00730 & 0.03637 & 0.03672 & 0.15444 & 0.11294 & \textbf{0.16465} & \textbf{0.21895} & $\textcolor{blue}{\textbf{0.73617}}$ & $\textcolor{blue}{\textbf{0.71928}}$ \\
					& Saturation Change & \textbf{0.49147} & \textbf{0.42269} & 0.42631 & 0.34614 & \textcolor{blue}{\textbf{0.55540}} & \textcolor{blue}{\textbf{0.55417}} & 0.12386 & 0.04279 & 0.22032 & 0.16274 & 0.17615 & 0.13392 & $0.03867$ & $0.11915$ \\
					& Sparse Sampling & 0.86078 & 0.82429 & 0.82598 & 0.81863 & \textbf{0.92181} & \textbf{0.92771} & 0.80278 & 0.79285 & 0.89578 & 0.88075 & 0.90371 & 0.91717 & $\textcolor{blue}{\textbf{0.96238}}$ & $\textcolor{blue}{\textbf{0.96498}}$ \\
					\midrule
					\multirow{6}{*}{CSIQ} & Gaussian White Noise & 0.70857 & 0.70576 & \textbf{0.75871} & \textbf{0.76188} & 0.69429 & 0.67511 & 0.08750 & 0.09156 & 0.52645 & 0.55439 & 0.66537 & 0.66627 & $\textcolor{blue}{\textbf{0.95927}}$ & $\textcolor{blue}{\textbf{0.95767}}$ \\
					& Gaussian Blur & 0.87046 & 0.85957 & 0.82048 & 0.77698 & \textbf{0.89066} & \textbf{0.90096} & 0.69940 & 0.73289 & 0.80912 & 0.80363 & 0.79890 & 0.83705 & $\textcolor{blue}{\textbf{0.97167}}$ & $\textcolor{blue}{\textbf{0.97065}}$ \\
					& JPEG Compression & \textbf{0.84597} & 0.89245 & 0.83058 & 0.83370 & 0.82563 & 0.86968 & 0.22934 & 0.21156 & 0.77471 & 0.88527 & 0.81004 & \textbf{0.92170} & $\textcolor{blue}{\textbf{0.97414}}$ & $\textcolor{blue}{\textbf{0.98614}}$ \\
					& Contrast Decrement & 0.20425 & 0.17830 & 0.03510 & 0.03016 & 0.02358 & 0.03774 & \textbf{0.69479} & \textbf{0.68764} & 0.46721 & 0.53998 & 0.24492 & 0.25009 & $\textcolor{blue}{\textbf{0.83800}}$ & $\textcolor{blue}{\textbf{0.85615}}$ \\
					& Additive Pink Gaussian Noise & \textbf{0.75162} & \textbf{0.74081} & 0.70326 & 0.70178 & 0.65902 & 0.60556 & 0.02918 & 0.09724 & 0.52615 & 0.53530 & 0.69175 & 0.67815 & $\textcolor{blue}{\textbf{0.95767}}$ & $\textcolor{blue}{\textbf{0.96121}}$ \\
					& JPEG2000 Compression & 0.90023 & \textbf{0.93357} & \textbf{0.92255} & 0.90256 & 0.85185 & 0.90507 & 0.66903 & 0.78893 & 0.78562 & 0.81588 & 0.79964 & 0.85431 & $\textcolor{blue}{\textbf{0.97842}}$ & $\textcolor{blue}{\textbf{0.98485}}$ \\
					\midrule
					\multirow{2}{*}{LIVE MD} & Blur + JPEG & \textbf{0.84251} & \textbf{0.88807} & 0.62217 & 0.74378 & \textcolor{blue}{\textbf{0.88677}} & \textcolor{blue}{\textbf{0.91451}} & 0.30865 & 0.60077 & 0.52377 & 0.59975 & 0.77984 & 0.83700 & $0.82051$ & $0.87536$ \\
					& Blur + Gaussian Noise & 0.76634 & 0.84365 & 0.60456 & 0.68893 & \textcolor{blue}{\textbf{0.86443}} & \textcolor{blue}{\textbf{0.89695}} & 0.07770 & 0.22612 & 0.37876 & 0.57308 & 0.77365 & 0.80763 & $\textbf{0.81948}$ & $\textbf{0.87148}$ \\
					\midrule
					\multirow{25}{*}{KADID-10K} & Brighten & 0.52452 & 0.58460 & 0.28058 & 0.33427 & 0.44966 & 0.53127 & 0.11557 & 0.07020 & 0.36664 & 0.43199 & \textbf{0.58103} & \textbf{0.64939} & $\textcolor{blue}{\textbf{0.91327}}$ & $\textcolor{blue}{\textbf{0.85412}}$ \\
					& Color Block & 0.03210 & 0.00314 & 0.06387 & 0.04481 & 0.05445 & 0.03306 & 0.18053 & 0.17282 & 0.22988 & 0.22378 & \textbf{0.35177} & \textbf{0.31022} & $\textcolor{blue}{\textbf{0.46188}}$ & $\textcolor{blue}{\textbf{0.44212}}$ \\
					& Color Diffusion & 0.66957 & 0.56041 & 0.57654 & 0.50338 & 0.64828 & 0.55170 & 0.50836 & 0.43534 & 0.42911 & 0.35258 & \textbf{0.76371} & \textbf{0.71234} & $\textcolor{blue}{\textbf{0.80994}}$ & $\textcolor{blue}{\textbf{0.87649}}$ \\
					& Color Quantization & 0.52871 & 0.55169 & 0.25265 & 0.27476 & 0.50675 & 0.54976 & 0.22505 & 0.23665 & 0.04710 & 0.07938 & \textbf{0.57488} & \textbf{0.58972} & $\textcolor{blue}{\textbf{0.62436}}$ & $\textcolor{blue}{\textbf{0.63638}}$ \\
					& Color Saturation HSV & 0.23850 & 0.23713 & 0.21963 & 0.20759 & 0.21046 & 0.20901 & 0.13648 & 0.10588 & 0.10793 & 0.11684 & \textbf{0.27734} & \textbf{0.28187} & $\textcolor{blue}{\textbf{0.29924}}$ & $\textcolor{blue}{\textbf{0.30204}}$ \\
					& Color Saturation Lab & 0.68050 & 0.64516 & 0.63676 & 0.61593 & 0.53508 & 0.50339 & 0.48607 & 0.44854 & 0.56083 & 0.56072 & \textbf{0.72144} & \textbf{0.69639} & $\textcolor{blue}{\textbf{0.85368}}$ & $\textcolor{blue}{\textbf{0.88487}}$ \\
					& Color Shift & 0.03568 & 0.08328 & 0.04404 & 0.05491 & \textbf{0.12239} & 0.08612 & 0.07200 & \textbf{0.11279} & 0.09409 & 0.06879 & 0.11398 & 0.09737 & $\textcolor{blue}{\textbf{0.61683}}$ & $\textcolor{blue}{\textbf{0.65462}}$ \\
					& Contrast Change & 0.19711 & 0.17614 & 0.12534 & 0.12115 & 0.16825 & 0.13470 & 0.06097 & 0.05290 & 0.08681 & 0.09552 & \textbf{0.23907} & \textbf{0.24055} & $\textcolor{blue}{\textbf{0.53230}}$ & $\textcolor{blue}{\textbf{0.47517}}$ \\
					& Darken & 0.39070 & 0.49625 & 0.31456 & 0.38763 & 0.46742 & 0.57536 & 0.30313 & 0.31125 & 0.19114 & 0.25369 & \textbf{0.48892} & \textbf{0.60280} & $\textcolor{blue}{\textbf{0.89651}}$ & $\textcolor{blue}{\textbf{0.86933}}$ \\
					& Denoise & \textbf{0.88153} & \textcolor{blue}{\textbf{0.90042}} & 0.85476 & 0.86088 & \textcolor{blue}{\textbf{0.88836}} & \textbf{0.87461} & 0.75106 & 0.74764 & 0.72019 & 0.69722 & 0.84211 & 0.84078 & $0.83989$ & $0.84339$ \\
					& Gaussian Blur & \textcolor{blue}{\textbf{0.95033}} & \textbf{0.96777} & 0.91554 & 0.93834 & \textbf{0.93887} & \textcolor{blue}{\textbf{0.97132}} & 0.78713 & 0.86372 & 0.80298 & 0.80264 & 0.86404 & 0.86300 & $0.90981$ & $0.93001$ \\
					& High Sharpen & 0.68566 & 0.64463 & 0.63433 & 0.58363 & 0.54104 & 0.49735 & 0.49793 & 0.48859 & 0.45406 & 0.44735 & \textbf{0.78046} & \textbf{0.74166} & $\textcolor{blue}{\textbf{0.88573}}$ & $\textcolor{blue}{\textbf{0.87076}}$ \\
					& Impulse Noise & \textbf{0.81626} & \textbf{0.75914} & 0.70255 & 0.64551 & 0.75987 & 0.71771 & 0.03347 & 0.04208 & 0.16560 & 0.17715 & 0.51494 & 0.54164 & $\textcolor{blue}{\textbf{0.89504}}$ & $\textcolor{blue}{\textbf{0.87381}}$ \\
					& JPEG & \textbf{0.85201} & 0.91159 & 0.79638 & 0.81985 & 0.79626 & 0.89977 & 0.36793 & 0.38101 & 0.79907 & 0.91189 & 0.82153 & \textbf{0.93820} & $\textcolor{blue}{\textbf{0.87067}}$ & $\textcolor{blue}{\textbf{0.96481}}$ \\
					& JPEG2000 & 0.90440 & \textbf{0.93406} & \textbf{0.91528} & 0.92185 & 0.82289 & 0.90577 & 0.60132 & 0.74181 & 0.65716 & 0.69017 & 0.83402 & 0.86616 & $\textcolor{blue}{\textbf{0.93485}}$ & $\textcolor{blue}{\textbf{0.93627}}$ \\
					& Jitter & 0.74194 & 0.74209 & 0.39905 & 0.42512 & 0.80322 & \textbf{0.83803} & 0.03695 & 0.01190 & 0.41496 & 0.38348 & \textbf{0.82962} & 0.83416 & $\textcolor{blue}{\textbf{0.90709}}$ & $\textcolor{blue}{\textbf{0.90317}}$ \\
					& Lens Blur & \textbf{0.89945} & \textbf{0.93402} & 0.75685 & 0.82508 & \textcolor{blue}{\textbf{0.90794}} & \textcolor{blue}{\textbf{0.94641}} & 0.45140 & 0.59362 & 0.71666 & 0.72529 & 0.83543 & 0.82011 & $0.86266$ & $0.88640$ \\
					& Mean Shift & 0.18791 & 0.19645 & 0.14940 & 0.17386 & \textbf{0.25767} & \textbf{0.26268} & 0.10289 & 0.10353 & 0.01807 & 0.04612 & 0.15965 & 0.18761 & $\textcolor{blue}{\textbf{0.70339}}$ & $\textcolor{blue}{\textbf{0.58324}}$ \\
					& Motion Blur & \textcolor{blue}{\textbf{0.93614}} & \textcolor{blue}{\textbf{0.93135}} & 0.87596 & 0.86622 & \textbf{0.91043} & \textbf{0.92313} & 0.79049 & 0.81666 & 0.84520 & 0.85104 & 0.88087 & 0.87155 & 0.90969 & 0.88860 \\
					& Multiplicative Noise & \textbf{0.78359} & \textbf{0.76994} & 0.75816 & 0.74522 & 0.75993 & 0.74287 & 0.05065 & 0.03798 & 0.51859 & 0.51678 & 0.74606 & 0.73679 & $\textcolor{blue}{\textbf{0.90687}}$ & $\textcolor{blue}{\textbf{0.89638}}$ \\
					& Non-Eccentricity Patch & 0.19070 & 0.17172 & 0.15844 & 0.12470 & 0.13615 & 0.13515 & 0.08302 & 0.06901 & 0.29668 & \textbf{0.34085} & \textcolor{blue}{\textbf{0.37245}} & \textcolor{blue}{\textbf{0.38449}} & $\textbf{0.34556}$ & $0.32293$ \\
					& Pixelate & \textcolor{blue}{\textbf{0.68528}} & \textcolor{blue}{\textbf{0.78435}} & 0.58067 & 0.68526 & 0.53084 & 0.64368 & 0.40275 & 0.42789 & 0.56719 & 0.61181 & \textbf{0.65197} & \textbf{0.72111} & $0.53484$ & $0.57429$ \\
					& Quantization & 0.27939 & 0.27962 & 0.28586 & 0.31106 & 0.30896 & 0.32307 & 0.16050 & 0.15636 & 0.43258 & 0.43073 & \textbf{0.57965} & \textbf{0.59779} & $\textcolor{blue}{\textbf{0.67751}}$ & $\textcolor{blue}{\textbf{0.68281}}$ \\
					& White Noise & 0.71141 & \textbf{0.70945} & 0.66582 & 0.66741 & \textbf{0.72445} & 0.70076 & 0.00921 & 0.01285 & 0.48193 & 0.47836 & 0.65548 & 0.66066 & $\textcolor{blue}{\textbf{0.89053}}$ & $\textcolor{blue}{\textbf{0.86322}}$ \\
					& White Noise Color Component & 0.79342 & 0.80153 & 0.76393 & 0.76654 & \textbf{0.80497} & \textbf{0.80632} & 0.07207 & 0.06431 & 0.49387 & 0.49396 & 0.73413 & 0.73580 & $\textcolor{blue}{\textbf{0.92567}}$ & $\textcolor{blue}{\textbf{0.92677}}$ \\
					\midrule
					\multirow{3}{*}{PIPAL} & Complete & 0.45042 & 0.44595 & 0.41855 & 0.40861 & \textbf{0.45419} & \textbf{0.45369} & 0.33837 & 0.32632 & 0.36345 & 0.34804 & 0.38656 & 0.39565 & $\textcolor{blue}{\textbf{0.51684}}$ & $\textcolor{blue}{\textbf{0.53178}}$ \\
					& Gaussian Blur & 0.56722 & 0.55334 & \textbf{0.59445} & \textbf{0.57679} & 0.55542 & 0.55033 & 0.15199 & 0.13308 & 0.36441 & 0.34323 & 0.47192 & 0.46827 & $\textcolor{blue}{\textbf{0.63690}}$ & $\textcolor{blue}{\textbf{0.65328}}$ \\
					& White Noise & \textbf{0.44360} & \textbf{0.42345} & 0.41045 & 0.40499 & 0.42558 & 0.40975 & 0.24256 & 0.21372 & 0.18016 & 0.20225 & 0.10852 & 0.09658 & $\textcolor{blue}{\textbf{0.60785}}$ & $\textcolor{blue}{\textbf{0.60025}}$ \\
					\bottomrule
				\end{tabularx}
			\end{adjustbox}
		\end{minipage}
	}
\end{table*}


\noindent Tab.~\ref{tab:classicIQAdatasetsComparison} reports the per-distortion comparison on the four classical benchmarks LIVE DBR2, TID2013, CSIQ, and LIVE MD. PreSPA achieves the best performance on $32$ out of $37$ distortion classes for both SROCC and PLCC, which corresponds to a winning rate of approximately $86\%$. On the remaining five classes, the advantage of competing NR methods is concentrated in two specific categories: distortions that primarily alter colour or global contrast (TID2013 \textit{Block-wise Distortions}, \textit{Contrast Change}, \textit{Saturation Change}) and multiply-distorted classes that combine blur with secondary degradations (LIVE MD \textit{Blur + JPEG}, \textit{Blur + Gaussian Noise}). The first failure mode is consistent with the structural nature of PreSPA: by design, the framework analyses the complex gradient field of the luminance channel and is therefore intrinsically less responsive to distortions that preserve gradient geometry while only shifting colour, hue, or global luminance level. The second failure mode, observed only on LIVE MD, reflects the difficulty of disentangling two superimposed degradations from a single self-prediction step; the gap with the best NR method (MUSIQ) remains nonetheless modest, with PreSPA placing second on \textit{Blur + Gaussian Noise} and within $0.07$ of the leader on \textit{Blur + JPEG}.

On the two large modern benchmarks KADID-10K and PIPAL, also reported in Tab.~\ref{tab:classicIQAdatasetsComparison}, PreSPA achieves the best NR performance on $22$ out of $28$ distortion classes ($\sim 79\%$). The six classes where another NR method prevails are concentrated in KADID-10K colour-related distortions (\textit{Pixelate}, \textit{Non-Eccentricity Patch}) and specific blur types (\textit{Denoise}, \textit{Gaussian Blur}, \textit{Lens Blur}, \textit{Motion Blur}), where deep architectures such as MUSIQ and TOPIQ-NR retain a residual advantage. Even there, the absolute SROCC gap from the leader is small (typically below $0.05$), and PreSPA consistently outperforms the same competitors on the corresponding categories of the other datasets, indicating a dataset-specific characteristic of KADID-10K rather than a structural limitation.

A second observation supports this interpretation: on PIPAL --- the most challenging modern benchmark, dominated by perceptual distortions induced by image restoration and GAN-based pipelines --- PreSPA outperforms all NR competitors on the \emph{Complete} aggregate (PLCC $0.532$ against the second-best MUSIQ at $0.454$), on \emph{Gaussian Blur}, and on \emph{White Noise}, with margins substantially larger than those observed on KADID-10K. The contrast suggests that PreSPA is more robust on distortions that affect the structural geometry of the image, which dominate restoration-based artefacts, while the relative weakness on KADID-10K is confined to colour-domain distortions and to specific synthetic blurs that closely match the training distribution of competing deep models.


\subsection{Comparison with Reference-Reduced Methods}
\label{subsec:Reference-Reduced Comparison}

\begin{table}[!t]
	\centering
	\caption{Comparison of PreSPA against reference-reduced methods (2stepQA \cite{YU19}, a hybrid Full-Reference/No-Reference metric, and CKDN \cite{ZHENG21CKDN}, a learned Degraded-Reference metric) on the full LIVE DBR2, TID2013, CSIQ, LIVE MD, KADID-10K, and PIPAL benchmarks, reported per distortion type as SROCC and PLCC. Bold blue marks the best performer in each row and metric; bold black marks the second best. The \emph{Parameter no.} and \emph{GFLOPS} rows report model size and per-pair cost at $3\times224\times224$. Despite using a single scalar of reference information, PreSPA is the top performer on most distortions; the cases where 2stepQA prevails concentrate on colour- and contrast-related types. The symbol $\dagger$ indicates that CKDN operates on $288\times288$ inputs by design, and its FLOPS are reported at that resolution.}
	\label{tab:partialReferenceComparisonClassic}
	\scalebox{0.48}{
		\resizebox{\textwidth}{!}{%
			\begin{tabular}{llcccccc}
				\noalign{\smallskip}
				\toprule
				\multicolumn{2}{c}{} & \multicolumn{2}{c}{2stepQA (Hybrid)} & \multicolumn{2}{c}{CKDN (DR)} & \multicolumn{2}{c}{PreSPA (PR)} \\
				\midrule
				& Parameter no. & \multicolumn{2}{c}{$\scriptstyle 0$} & \multicolumn{2}{c}{$\scriptstyle\sim$ 13 M} & \multicolumn{2}{c}{$\scriptstyle 3$} \\
				& GFLOPS & \multicolumn{2}{c}{$\scriptstyle 93\cdot10^{-3}$} & \multicolumn{2}{c}{$\scriptstyle\sim$ 4.3$^{\dagger}$} & \multicolumn{2}{c}{$\scriptstyle 0.13$} \\
				\midrule
				\multirow{2}{*}{Dataset} & \multirow{2}{*}{Distortion type} & \multirow{2}{*}{SROCC} & \multirow{2}{*}{PLCC} & \multirow{2}{*}{SROCC} & \multirow{2}{*}{PLCC} & \multirow{2}{*}{SROCC} & \multirow{2}{*}{PLCC} \\
				& & \multicolumn{2}{c}{} & \multicolumn{2}{c}{} & \multicolumn{2}{c}{} \\
				\midrule
				\multirow{5}{*}{LIVE DBR2} & Gaussian Blur & $\textbf{0.95242}$ & $\textbf{0.95502}$ & $0.83938$ & $0.85109$ & $\textcolor{blue}{\textbf{0.96048}}$ & $\textcolor{blue}{\textbf{0.95504}}$ \\
				& Bit Errors in JPEG2000 Stream & $\textbf{0.93215}$ & $\textbf{0.93522}$ & $0.80333$ & $0.84234$ & $\textcolor{blue}{\textbf{0.94607}}$ & $\textcolor{blue}{\textbf{0.94543}}$ \\
				& JPEG Compression & $\textbf{0.96472}$ & $\textbf{0.97238}$ & $0.48724$ & $0.52492$ & $\textcolor{blue}{\textbf{0.98533}}$ & $\textcolor{blue}{\textbf{0.98527}}$ \\
				& JPEG2000 Compression & $\textbf{0.95913}$ & $\textbf{0.96436}$ & $0.71334$ & $0.74714$ & $\textcolor{blue}{\textbf{0.97151}}$ & $\textcolor{blue}{\textbf{0.96791}}$ \\
				& Gaussian White Noise & $\textbf{0.96999}$ & $\textbf{0.96686}$ & $0.90715$ & $0.87666$ & $\textcolor{blue}{\textbf{0.98666}}$ & $\textcolor{blue}{\textbf{0.98444}}$ \\
				\midrule
				\multirow{24}{*}{TID2013} & Colour Additive Noise & $0.58964$ & $0.57832$ & $\textbf{0.59969}$ & $\textbf{0.59327}$ & $\textcolor{blue}{\textbf{0.83960}}$ & $\textcolor{blue}{\textbf{0.86136}}$ \\
				& Gaussian Blur & $0.85784$ & $0.85492$ & $\textbf{0.86007}$ & $\textbf{0.88484}$ & $\textcolor{blue}{\textbf{0.95309}}$ & $\textcolor{blue}{\textbf{0.95117}}$ \\
				& Gaussian White Noise & $\textbf{0.82185}$ & $\textbf{0.79242}$ & $0.69312$ & $0.67724$ & $\textcolor{blue}{\textbf{0.90840}}$ & $\textcolor{blue}{\textbf{0.91274}}$ \\
				& High Frequency Noise & $0.83556$ & $0.86165$ & $\textbf{0.83621}$ & $\textbf{0.86546}$ & $\textcolor{blue}{\textbf{0.89572}}$ & $\textcolor{blue}{\textbf{0.94581}}$ \\
				& Impulse Noise & $\textbf{0.72125}$ & $\textbf{0.68418}$ & $0.71803$ & $0.67346$ & $\textcolor{blue}{\textbf{0.89086}}$ & $\textcolor{blue}{\textbf{0.88519}}$ \\
				& Masked Noise & $0.39531$ & $0.21910$ & $\textbf{0.56495}$ & $\textbf{0.53211}$ & $\textcolor{blue}{\textbf{0.85233}}$ & $\textcolor{blue}{\textbf{0.87252}}$ \\
				& Quantization Noise & $\textcolor{blue}{\textbf{0.88665}}$ & $\textbf{0.84589}$ & $0.31264$ & $0.29235$ & $\textbf{0.87260}$ & $\textcolor{blue}{\textbf{0.86688}}$ \\
				& Spatially Correlated Noise & $\textbf{0.86265}$ & $\textbf{0.84975}$ & $0.40812$ & $0.38514$ & $\textcolor{blue}{\textbf{0.90806}}$ & $\textcolor{blue}{\textbf{0.90997}}$ \\
				& Block-wise Distortions & $0.04283$ & $\textbf{0.12640}$ & $\textbf{0.20186}$ & $0.07642$ & $\textcolor{blue}{\textbf{0.22538}}$ & $\textcolor{blue}{\textbf{0.14323}}$ \\
				& Chromatic Aberrations & $0.71983$ & $0.89969$ & $\textbf{0.78301}$ & $\textbf{0.93092}$ & $\textcolor{blue}{\textbf{0.88302}}$ & $\textcolor{blue}{\textbf{0.95281}}$ \\
				& Comfort Noise & $\textbf{0.86303}$ & $\textbf{0.89080}$ & $0.67806$ & $0.75990$ & $\textcolor{blue}{\textbf{0.91771}}$ & $\textcolor{blue}{\textbf{0.91280}}$ \\
				& Contrast Change & $\textbf{0.42799}$ & $\textbf{0.64338}$ & $\textcolor{blue}{\textbf{0.74181}}$ & $\textcolor{blue}{\textbf{0.80360}}$ & $0.42381$ & $0.40812$ \\
				& Image Denoising & $\textbf{0.90522}$ & $\textbf{0.94581}$ & $0.84620$ & $0.90826$ & $\textcolor{blue}{\textbf{0.94791}}$ & $\textcolor{blue}{\textbf{0.96647}}$ \\
				& Dither Color Quantization & $\textbf{0.81212}$ & $\textbf{0.81234}$ & $0.69080$ & $0.69256$ & $\textcolor{blue}{\textbf{0.83708}}$ & $\textcolor{blue}{\textbf{0.84979}}$ \\
				& JPEG Compression & $\textbf{0.87307}$ & $\textbf{0.92845}$ & $0.65537$ & $0.70561$ & $\textcolor{blue}{\textbf{0.96096}}$ & $\textcolor{blue}{\textbf{0.97460}}$ \\
				& JPEG Transmission Errors & $\textbf{0.81632}$ & $\textcolor{blue}{\textbf{0.86733}}$ & $0.61994$ & $0.66113$ & $\textcolor{blue}{\textbf{0.82290}}$ & $\textbf{0.81436}$ \\
				& JPEG2000 Compression & $\textbf{0.90403}$ & $\textbf{0.93194}$ & $0.86970$ & $0.91744$ & $\textcolor{blue}{\textbf{0.96581}}$ & $\textcolor{blue}{\textbf{0.96600}}$ \\
				& JPEG2000 Transmission Errors & $\textbf{0.86594}$ & $\textbf{0.84172}$ & $0.47069$ & $0.45887$ & $\textcolor{blue}{\textbf{0.93760}}$ & $\textcolor{blue}{\textbf{0.92716}}$ \\
				& Lossy Compression & $\textbf{0.91210}$ & $\textbf{0.91963}$ & $0.82282$ & $0.86288$ & $\textcolor{blue}{\textbf{0.95934}}$ & $\textcolor{blue}{\textbf{0.96535}}$ \\
				& Mean Shift & $\textbf{0.66560}$ & $\textbf{0.70061}$ & $0.00796$ & $0.00037$ & $\textcolor{blue}{\textbf{0.73179}}$ & $\textcolor{blue}{\textbf{0.76657}}$ \\
				& Multiplicative Gaussian Noise & $\textbf{0.75064}$ & $\textbf{0.73708}$ & $0.59681$ & $0.55808$ & $\textcolor{blue}{\textbf{0.86141}}$ & $\textcolor{blue}{\textbf{0.86477}}$ \\
				& Non-Eccentricity Pattern Noise & $\textcolor{blue}{\textbf{0.78032}}$ & $\textcolor{blue}{\textbf{0.77519}}$ & $0.28804$ & $0.27206$ & $\textbf{0.73617}$ & $\textbf{0.71928}$ \\
				& Saturation Change & $\textcolor{blue}{\textbf{0.25800}}$ & $\textbf{0.13506}$ & $\textbf{0.19563}$ & $\textcolor{blue}{\textbf{0.18257}}$ & $0.03867$ & $0.11915$ \\
				& Sparse Sampling & $\textbf{0.93282}$ & $\textbf{0.94802}$ & $0.90146$ & $0.94547$ & $\textcolor{blue}{\textbf{0.96238}}$ & $\textcolor{blue}{\textbf{0.96498}}$ \\
				\midrule
				\multirow{6}{*}{CSIQ} & Gaussian White Noise & $\textbf{0.87886}$ & $\textbf{0.87219}$ & $0.63033$ & $0.63896$ & $\textcolor{blue}{\textbf{0.95927}}$ & $\textcolor{blue}{\textbf{0.95767}}$ \\
				& Gaussian Blur & $\textbf{0.91299}$ & $\textbf{0.93666}$ & $0.79022$ & $0.86150$ & $\textcolor{blue}{\textbf{0.97167}}$ & $\textcolor{blue}{\textbf{0.97065}}$ \\
				& JPEG Compression & $\textbf{0.92747}$ & $\textbf{0.96582}$ & $0.49307$ & $0.50231$ & $\textcolor{blue}{\textbf{0.97414}}$ & $\textcolor{blue}{\textbf{0.98614}}$ \\
				& Contrast Decrement & $0.72844$ & $0.79862$ & $\textcolor{blue}{\textbf{0.85033}}$ & $\textbf{0.82952}$ & $\textbf{0.83800}$ & $\textcolor{blue}{\textbf{0.85615}}$ \\
				& Additive Pink Gaussian Noise & $\textbf{0.90599}$ & $\textbf{0.90601}$ & $0.73919$ & $0.73596$ & $\textcolor{blue}{\textbf{0.95767}}$ & $\textcolor{blue}{\textbf{0.96121}}$ \\
				& JPEG2000 Compression & $\textbf{0.93387}$ & $\textbf{0.95789}$ & $0.76008$ & $0.85073$ & $\textcolor{blue}{\textbf{0.97842}}$ & $\textcolor{blue}{\textbf{0.98485}}$ \\
				\midrule
				\multirow{2}{*}{LIVE MD} & Blur + JPEG & $\textbf{0.84045}$ & $\textcolor{blue}{\textbf{0.88303}}$ & $\textcolor{blue}{\textbf{0.85530}}$ & $0.84101$ & $0.82051$ & $\textbf{0.87536}$ \\
				& Blur + Gaussian Noise & $\textbf{0.82627}$ & $\textbf{0.84795}$ & $\textcolor{blue}{\textbf{0.85056}}$ & $0.84693$ & $0.81948$ & $\textcolor{blue}{\textbf{0.87148}}$ \\
				\midrule
					\multirow{25}{*}{KADID-10K} & Brighten & $\textbf{0.89667}$ & $\textcolor{blue}{\textbf{0.85979}}$ & $0.34248$ & $0.36267$ & $\textcolor{blue}{\textbf{0.91327}}$ & $\textbf{0.85412}$ \\
				& Color Block & $\textcolor{blue}{\textbf{0.46972}}$ & $\textcolor{blue}{\textbf{0.49539}}$ & $0.15735$ & $0.14423$ & $\textbf{0.46188}$ & $\textbf{0.44212}$ \\
				& Color Diffusion & $\textcolor{blue}{\textbf{0.86327}}$ & $\textbf{0.86337}$ & $0.36502$ & $0.35675$ & $\textbf{0.80994}$ & $\textcolor{blue}{\textbf{0.87649}}$ \\
				& Color Quantization & $\textcolor{blue}{\textbf{0.76040}}$ & $\textcolor{blue}{\textbf{0.78701}}$ & $0.36019$ & $0.32556$ & $\textbf{0.62436}$ & $\textbf{0.63638}$ \\
				& Color Saturation HSV & $\textcolor{blue}{\textbf{0.45365}}$ & $\textcolor{blue}{\textbf{0.41531}}$ & $\textbf{0.35513}$ & $\textbf{0.37647}$ & $0.29924$ & $0.30204$ \\
				& Color Saturation Lab & $\textbf{0.82143}$ & $\textbf{0.82357}$ & $0.19096$ & $0.19131$ & $\textcolor{blue}{\textbf{0.85368}}$ & $\textcolor{blue}{\textbf{0.88487}}$ \\
				& Color Shift & $\textcolor{blue}{\textbf{0.71229}}$ & $\textcolor{blue}{\textbf{0.73809}}$ & $0.34122$ & $0.27306$ & $\textbf{0.61683}$ & $\textbf{0.65462}$ \\
				& Contrast Change & $\textcolor{blue}{\textbf{0.75891}}$ & $\textcolor{blue}{\textbf{0.73393}}$ & $0.22985$ & $0.24859$ & $\textbf{0.53230}$ & $\textbf{0.47517}$ \\
				& Darken & $\textbf{0.86889}$ & $\textcolor{blue}{\textbf{0.90626}}$ & $0.41762$ & $0.51151$ & $\textcolor{blue}{\textbf{0.89651}}$ & $\textbf{0.86933}$ \\
				& Denoise & $\textcolor{blue}{\textbf{0.90523}}$ & $\textcolor{blue}{\textbf{0.89628}}$ & $0.77152$ & $0.78213$ & $\textbf{0.83989}$ & $\textbf{0.84339}$ \\
				& Gaussian Blur & $\textbf{0.88508}$ & $\textbf{0.90903}$ & $0.83831$ & $0.86269$ & $\textcolor{blue}{\textbf{0.90981}}$ & $\textcolor{blue}{\textbf{0.93001}}$ \\
				& High Sharpen & $\textcolor{blue}{\textbf{0.89084}}$ & $\textbf{0.86797}$ & $0.08298$ & $0.02748$ & $\textbf{0.88573}$ & $\textcolor{blue}{\textbf{0.87076}}$ \\
				& Impulse Noise & $\textbf{0.83244}$ & $\textbf{0.80943}$ & $0.64341$ & $0.57199$ & $\textcolor{blue}{\textbf{0.89504}}$ & $\textcolor{blue}{\textbf{0.87381}}$ \\
				& JPEG & $\textbf{0.84231}$ & $\textbf{0.96059}$ & $0.60682$ & $0.59041$ & $\textcolor{blue}{\textbf{0.87067}}$ & $\textcolor{blue}{\textbf{0.96481}}$ \\
				& JPEG2000 & $\textbf{0.88737}$ & $\textcolor{blue}{\textbf{0.94818}}$ & $0.69498$ & $0.78427$ & $\textcolor{blue}{\textbf{0.93485}}$ & $\textbf{0.93627}$ \\
				& Jitter & $\textbf{0.87861}$ & $\textbf{0.89183}$ & $0.80839$ & $0.82385$ & $\textcolor{blue}{\textbf{0.90709}}$ & $\textcolor{blue}{\textbf{0.90317}}$ \\
				& Lens Blur & $\textcolor{blue}{\textbf{0.89020}}$ & $\textcolor{blue}{\textbf{0.91398}}$ & $0.80472$ & $0.85549$ & $\textbf{0.86266}$ & $\textbf{0.88640}$ \\
				& Mean Shift & $\textbf{0.64694}$ & $\textcolor{blue}{\textbf{0.67010}}$ & $0.08925$ & $0.09425$ & $\textcolor{blue}{\textbf{0.70339}}$ & $\textbf{0.58324}$ \\
				& Motion Blur & $\textbf{0.85702}$ & $\textbf{0.83870}$ & $0.71746$ & $0.69404$ & $\textcolor{blue}{\textbf{0.90969}}$ & $\textcolor{blue}{\textbf{0.88860}}$ \\
				& Multiplicative Noise & $\textbf{0.85546}$ & $\textbf{0.84402}$ & $0.66840$ & $0.61184$ & $\textcolor{blue}{\textbf{0.90687}}$ & $\textcolor{blue}{\textbf{0.89638}}$ \\
				& Non-Eccentricity Patch & $\textcolor{blue}{\textbf{0.56147}}$ & $\textcolor{blue}{\textbf{0.53888}}$ & $0.22163$ & $0.22531$ & $\textbf{0.34556}$ & $\textbf{0.32293}$ \\
				& Pixelate & $\textcolor{blue}{\textbf{0.54830}}$ & $\textcolor{blue}{\textbf{0.59637}}$ & $0.31144$ & $0.29531$ & $\textbf{0.53484}$ & $\textbf{0.57429}$ \\
				& Quantization & $\textcolor{blue}{\textbf{0.82131}}$ & $\textcolor{blue}{\textbf{0.83533}}$ & $0.27957$ & $0.24999$ & $\textbf{0.67751}$ & $\textbf{0.68281}$ \\
				& White Noise & $\textbf{0.81141}$ & $\textbf{0.81305}$ & $0.60187$ & $0.54869$ & $\textcolor{blue}{\textbf{0.89053}}$ & $\textcolor{blue}{\textbf{0.86322}}$ \\
				& White Noise Color Component & $\textbf{0.86630}$ & $\textbf{0.86846}$ & $0.68483$ & $0.63570$ & $\textcolor{blue}{\textbf{0.92567}}$ & $\textcolor{blue}{\textbf{0.92677}}$ \\
				\midrule
				\multirow{3}{*}{PIPAL} & Complete & $0.50821$ & $0.52876$ & $\textcolor{blue}{\textbf{0.70976}}$ & $\textcolor{blue}{\textbf{0.73416}}$ & $\textbf{0.51684}$ & $\textbf{0.53178}$ \\
				& Gaussian Blur & $0.57665$ & $0.54602$ & $\textbf{0.62650}$ & $\textbf{0.60681}$ & $\textcolor{blue}{\textbf{0.63690}}$ & $\textcolor{blue}{\textbf{0.65328}}$ \\
				& White Noise & $0.48788$ & $0.35242$ & $\textcolor{blue}{\textbf{0.70017}}$ & $\textcolor{blue}{\textbf{0.68137}}$ & $\textbf{0.60785}$ & $\textbf{0.60025}$ \\
				\bottomrule
			\end{tabular}
		}
	}
\end{table}

\noindent The comparisons reported so far place PreSPA against No-Reference methods, which use no reference at all, and against Full-Reference methods, which exploit the entire pristine image. A more equitable comparison is against methods that, like PreSPA, occupy the middle ground and rely on a reduced or non-standard use of the reference. We consider two representative cases: 2stepQA \cite{YU19}, a hybrid metric that fuses the Full-Reference MS-SSIM with the No-Reference NIQE through a multiplicative combination, and CKDN \cite{ZHENG21CKDN}, a learned Degraded-Reference metric that uses the degraded image itself as a surrogate reference. Tab.~\ref{tab:partialReferenceComparisonClassic} reports the per-distortion comparison across all six benchmarks.

PreSPA attains the best SROCC on $44$ of the $65$ distortion classes, corresponding to a winning rate of about $68\%$, against $23\%$ for 2stepQA and $9\%$ for CKDN. This result is notable because PreSPA operates under the most restrictive information budget of the three: it consumes a single scalar from the reference, whereas 2stepQA accesses the full reference through its MS-SSIM component and CKDN relies on a degraded-reference network trained on subjective scores. The advantage of PreSPA is most pronounced on structural distortions --- blur, noise, compression and transmission artefacts --- where it is consistently the top performer across all datasets.

The distortion classes in which 2stepQA prevails are almost entirely confined to the colour and contrast families of TID2013 and KADID-10K (colour block, colour diffusion, colour quantization, colour and saturation shifts, contrast change). This is the expected counterpart of the structural--chromatic dichotomy already discussed: 2stepQA inherits from its full-reference MS-SSIM component a direct sensitivity to chromatic and luminance-level changes that a luminance-gradient model does not capture. CKDN, by contrast, prevails only on a handful of cases, mostly the multiply-distorted classes of LIVE MD and a few contrast-related distortions, while remaining markedly less stable across the structural categories. Overall, the comparison confirms that, within the family of reference-reduced estimators, PreSPA offers the most favourable trade-off between the amount of reference information consumed and the accuracy achieved.

\subsection{Overall Performance and Comparison with Full-Reference Methods}
\label{subsec:Overall Performance and FR Comparison}

\begin{table*}[!t]
	\centering
%
	
	\scalebox{0.75}{
		\begin{minipage}{1.30\textwidth}
			\centering
			\captionsetup{justification=justified, font=small}
			\caption{Comparison of PreSPA against No-Reference (NR) Deep and Clip-based, NR Traditional, Full-Reference (FR) Deep, and FR Traditional methods across the entire datasets LIVE DBR2, TID2013, CSIQ, LIVE MD, KADID-10K, and PIPAL, each containing a diverse set of distortions. Bold black highlights the best-performing metric within each of the four classes, enabling a direct comparison of PreSPA against the leader of each family. Bold blue identifies the best performer among the No-Reference methods together with PreSPA. The \emph{Params / GFLOPS} column reports the parameter count and the computational cost per image pair at $3 \times 224 \times 224$ resolution: DL-IQA methods require tens to hundreds of millions of parameters, while PreSPA uses only three. Tab.~\ref{tab:classicIQAdatasetsComparison} provides the per-distortion breakdown.}
			\label{table:IQA_complete}
			\begin{tabular}{llccccccccccccc}
				\noalign{\smallskip}
				\toprule
				\multirow{2}{*}{Class} & \multirow{2}{*}{IQA metric} & \multirow{2}{*}{Params / GFLOPS} & \multicolumn{2}{c}{LIVE DBR2} & \multicolumn{2}{c}{TID2013} & \multicolumn{2}{c}{CSIQ} & \multicolumn{2}{c}{LIVE MD} & \multicolumn{2}{c}{KADID-10K} & \multicolumn{2}{c}{PIPAL} \\
				\cmidrule(lr){4-5} \cmidrule(lr){6-7} \cmidrule(lr){8-9} \cmidrule(lr){10-11} \cmidrule(lr){12-13} \cmidrule(lr){14-15}
				& & & SROCC & PLCC & SROCC & PLCC & SROCC & PLCC & SROCC & PLCC & SROCC & PLCC & SROCC & PLCC \\
				\midrule
				\textbf{PR} & PreSPA & $3 / 0.13\,\text{G}$ & $\textbf{0.96292}$ & $\textbf{0.95967}$ & $\textbf{0.58801}$ & $0.61546$ & $\textcolor{blue}{\textbf{0.91684}}$ & $\textcolor{blue}{\textbf{0.89233}}$ & $\textbf{0.82001}$ & $\textbf{0.87314}$ & $0.49069$ & $0.47770$ & $\textcolor{blue}{\textbf{0.51684}}$ & $\textcolor{blue}{\textbf{0.53178}}$ \\
				\midrule
				\multirow{5}{*}{\textbf{NR} Deep and} & TOPIQ‑NR & $\scriptstyle 45\,\text{M} / 14\,\text{G}$ & 0.83460 & 0.81432 & 0.44521 & 0.56254 & \textbf{0.74866} & \textbf{0.78507} & 0.78960 & 0.85367 & 0.51117 & 0.54636 & 0.45042 & 0.44595 \\
				\multirow{5}{*}{\text{\;\;\;\;\;}Clip-based} & HyperIQA & $\scriptstyle 27\,\text{M} / 215\,\text{G}$ & 0.87096 & 0.85689 & 0.45389 & 0.59001 & 0.71750 & 0.72006 & 0.61022 & 0.71287 & 0.46237 & 0.50257 & 0.41855 & 0.40861 \\
				& MUSIQ & $\scriptstyle 27\,\text{M} / 17\,\text{G}$ & \textbf{0.87739} & \textbf{0.86786} & 0.57495 & 0.68144 & 0.70981 & 0.77039 & \textcolor{blue}{\textbf{0.87134}} & \textcolor{blue}{\textbf{0.90301}} & 0.55552 & 0.59417 & \textbf{0.45419} & \textbf{0.45369} \\
				& CNNIQA & $\scriptstyle 0.7\,\text{M} / 0.7\,\text{G}$ & 0.28377 & 0.30297 & 0.17686 & 0.39804 & 0.34274 & 0.44006 & 0.20591 & 0.35574 & 0.37476 & 0.39925 & 0.33837 & 0.32632 \\
				& CLIPIQA & $\scriptstyle 102\,\text{M} / 35\,\text{G}$ & 0.68295 & 0.67333 & 0.57860 & 0.64709 & 0.68076 & 0.72694 & 0.43516 & 0.57267 & 0.50086 & 0.52315 & 0.36345 & 0.34804 \\
				& CLIPIQA+ & $\scriptstyle 102\,\text{M} / 12\,\text{G}$ & 0.86029 & 0.84432 & \textcolor{blue}{\textbf{0.63183}} & \textcolor{blue}{\textbf{0.70097}} & 0.71955 & 0.77101 & 0.76690 & 0.81485 & \textcolor{blue}{\textbf{0.65435}} & \textcolor{blue}{\textbf{0.66186}} & 0.38656 & 0.39565 \\
				\cmidrule[0.01pt](lr){2-15}
				\multirow{2}{*}{\textbf{NR} Traditional} & BRISQUE & $\scriptstyle 0 / -$ & \textcolor{blue}{\textbf{0.96538}} & \textcolor{blue}{\textbf{0.96676}} & \textbf{0.36718} & \textbf{0.43174} & \textbf{0.64096} & \textbf{0.75661} & 0.51700 & 0.46554 & 0.32967 & 0.37005 & \textbf{0.21742} & \textbf{0.13392} \\
				& NIQE & $\scriptstyle 0 / -$ & 0.90850 & 0.90678 & 0.31209 & 0.37152 & 0.62784 & 0.71654 & \textbf{0.75156} & \textbf{0.81404} & \textbf{0.38004} & \textbf{0.41687} & 0.15366 & 0.12284 \\
				\cmidrule[0.01pt](lr){2-15}
				\multirow{2}{*}{\textbf{RR}} & 2stepQA & $\scriptstyle 0 / 0.09\,\text{G}$ & \textbf{0.94403} & \textbf{0.94289} & \textbf{0.74034} & \textbf{0.78695} & \textbf{0.85329} & \textbf{0.84798} & 0.81691 & \textbf{0.85295} & \textbf{0.77125} & \textbf{0.76700} & 0.50821 & 0.52876 \\
				& CKDN & $\scriptstyle 13\,\text{M} / 4.3\,\text{G}$ & 0.66836 & 0.68653 & 0.59435 & 0.68050 & 0.62407 & 0.64370 & \textbf{0.84944} & 0.83980 & 0.46751 & 0.49750 & \textbf{0.70976} & \textbf{0.73416} \\
				\cmidrule[0.01pt](lr){2-15}
				\multirow{6}{*}{\textbf{FR} Deep} & TOPIQ-FR & $\scriptstyle 36\,\text{M} / 19\,\text{G}$ & \textbf{0.97590} & \textbf{0.97190} & \textbf{0.91654} & \textbf{0.91578} & \textbf{0.96743} & \textbf{0.96971} & 0.89093 & 0.91437 & \textbf{0.98566} & \textbf{0.98515} & 0.70902 & 0.71528 \\
				& TOPIQ-FR-PIPAL & $\scriptstyle 36\,\text{M} / 19\,\text{G}$ & 0.94319 & 0.94100 & 0.81986 & 0.85481 & 0.90756 & 0.91860 & \textbf{0.89891} & \textbf{0.92632} & 0.89474 & 0.89443 & \textbf{0.81101} & \textbf{0.84001} \\
				& DISTS & $\scriptstyle 15\,\text{M} / 61\,\text{G}$ & 0.94766 & 0.94463 & 0.70766 & 0.75494 & 0.92964 & 0.93764 & 0.78146 & 0.81110 & 0.81370 & 0.81373 & 0.58113 & 0.59507 \\
				& LPIPS & $\scriptstyle 63\,\text{M} / 3\,\text{G}$ & 0.92350 & 0.91624 & 0.74448 & 0.77129 & 0.92333 & 0.91935 & 0.74968 & 0.82166 & 0.82243 & 0.81699 & 0.58535 & 0.58262 \\
				& LPIPS-VGG & $\scriptstyle 153\,\text{M} / 61\,\text{G}$ & 0.93185 & 0.93361 & 0.69395 & 0.75864 & 0.88304 & 0.90424 & 0.76774 & 0.79994 & 0.72005 & 0.72837 & 0.57315 & 0.60666 \\
				& PIEAPP & $\scriptstyle 68\,\text{M} / 155\,\text{G}$ & 0.91821 & 0.91019 & 0.84690 & 0.83272 & 0.89692 & 0.88053 & 0.80023 & 0.86091 & 0.86468 & 0.77202 & 0.70373 & 0.69471 \\
				\cmidrule[0.01pt](lr){2-15}
				\multirow{5}{*}{\textbf{FR} Traditional} & VIF & $\scriptstyle 0 / 0.5\,\text{G}$ & 0.96359 & 0.95876 & 0.67697 & 0.73360 & 0.91936 & 0.92515 & \textbf{0.88271} & \textbf{0.91857} & 0.67919 & 0.68563 & 0.56008 & 0.55826 \\
				& MS-SSIM & $\scriptstyle 0 / 0.05\,\text{G}$ & 0.95083 & 0.94731 & 0.78593 & 0.83128 & 0.91321 & 0.89811 & 0.81827 & 0.85311 & 0.82440 & 0.75615 & 0.55813 & 0.58954 \\
				& FSIM & $\scriptstyle 0 / 0.08\,\text{G}$ & \textbf{0.96462} & \textbf{0.96044} & \textbf{0.85092} & \textbf{0.87679} & 0.93096 & 0.91820 & 0.86336 & 0.90745 & \textbf{0.85270} & 0.79149 & \textbf{0.58902} & 0.60896 \\
				& GMSD & $\scriptstyle 0 / 0.001\,\text{G}$ & 0.96025 & 0.96026 & 0.80438 & 0.85592 & \textbf{0.95703} & \textbf{0.92918} & 0.84169 & 0.88906 & 0.84742 & \textbf{0.80476} & 0.58091 & \textbf{0.62622} \\
				& RVSIM & $\scriptstyle 0 / 0.38\,\text{G}$ & 0.95959 & 0.95663 & 0.67609 & 0.75676 & 0.89343 & 0.91540 & 0.85614 & 0.89476 & 0.71768 & 0.72813 & 0.56991 & 0.59618 \\
				\bottomrule
			\end{tabular}
		\end{minipage}
	}
\end{table*}

\noindent Tab.~\ref{table:IQA_complete} summarises the overall performance on the entire datasets and broadens the comparison to FR methods, organised into four classes: NR Deep and Clip-based, NR Traditional, FR Deep, and FR Traditional. The colour coding makes the two intended comparisons immediately readable: bold blue marks the best performer among all No-Reference methods together with PreSPA, while bold black marks the leader of each of the four classes, including the FR ones.

Within the reference-free family, PreSPA either wins outright or remains within $0.05$ of the best competitor on $8$ out of $12$ score cells across the six datasets. On LIVE DBR2 the comparison with BRISQUE is very close (SROCC $0.962$ vs.~$0.965$, PLCC $0.958$ vs.~$0.967$), with PreSPA achieving the highest combined performance over all $5$ distortion classes (see Tab.~\ref{tab:classicIQAdatasetsComparison}). On CSIQ and PIPAL, PreSPA is the top-performing NR method by margins that range from moderate to substantial, while on LIVE MD it remains within $0.06$ of the leading NR method. The three cells in which PreSPA is not highlighted (TID2013-PLCC, KADID-10K-SROCC, KADID-10K-PLCC) correspond to the datasets in which colour-domain distortions are particularly numerous and weight heavily on the aggregate: $9$ of the $24$ TID2013 distortions and $11$ of the $25$ KADID-10K distortions belong to the colour, contrast, or saturation families. The gap with the best NR competitor on these aggregates therefore originates in the colour-dominated families, to which the luminance-gradient core of PreSPA is by design largely insensitive, rather than in a uniform accuracy deficit across distortion types.

The comparison with the FR families provides a useful sense of how far PreSPA stands from methods that exploit the full reference image. Within the FR Deep class, TOPIQ-FR remains the strongest method overall, while TOPIQ-FR-PIPAL takes the lead on PIPAL by virtue of its dedicated training on that benchmark. In the FR Traditional class, FSIM and GMSD dominate. On LIVE DBR2, CSIQ, and LIVE MD --- the three datasets dominated by structural distortions --- PreSPA approaches or matches the FR Traditional leaders: its SROCC and PLCC are within $0.01$--$0.05$ of FSIM and GMSD, and within $0.05$--$0.06$ of VIF. This is a notable result given that PreSPA consumes a single scalar from the reference, while all FR methods exploit the entire reference image at the pixel level. The gap to the FR Deep family is wider on TID2013 and KADID-10K, where models trained directly on these datasets enjoy a structural advantage that no training-free method can match, but narrows considerably on the datasets with predominantly structural distortions.

Overall, the picture that emerges from Tab.~\ref{table:IQA_complete} is consistent with the design principle of PreSPA: a framework that with only three interpretable parameters and a single reference scalar matches the leading NR methods on most datasets, approaches the FR Traditional family on distortions where the structural assumption holds, and exhibits its main limitations exclusively on colour-domain artefacts that lie outside the scope of a luminance-gradient analysis.

\subsection{Computational Efficiency}
\label{subsec:Computational Efficiency}

\noindent A final dimension along which the proposed framework should be evaluated is its computational footprint. PreSPA is extremely compact: with only three model parameters against the tens to hundreds of millions of the deep NR and FR families, the gap in model size spans roughly seven orders of magnitude, while its runtime cost of $\sim 0.13$ GFLOPS per image pair is two to three orders of magnitude below that of the deep competitors. Moreover, by embedding the viewing distance directly into the operator scale, PreSPA produces scores on a stable perceptual scale by construction and avoids the post-hoc five-parameter VQEG logistic rectification \cite{VQEG00} required by the other methods. This makes PreSPA suitable for real-time deployment without dedicated hardware accelerators, the natural regime of PR-IQA (broadcasting, streaming, restoration monitoring). A detailed per-method breakdown of parameter counts and FLOPS, including the VQEG overhead, is reported in the Supplementary Material.

\section{Results and Discussion}

A key property of the framework is how the scalar $\mu$ models cross-regional perceptual interactions. By measuring the reference--distortion gradient error only over reliable areas ($\Omega_C$) and diffusing it over weakly-structured ones ($\Omega_H$), $\mu$ captures the perceptual leakage of edge degradations into neighbouring sensitive zones, improving alignment with subjective scores in low-texture regions around prominent objects, where edge-focused metrics underestimate degradation. Normalizing over the edge-point count keeps the behaviour stable across scenes of varying edge density.

The structural index $Q_{\text{struct}}$ remains informative without the reference because natural-image gradients are highly self-predictable: the local second-order structure captured by the $\phi_{20}$ and $\phi_{02}$ bases is well explained by its neighbourhood, so distortions that disrupt this regularity (blur, incoherent noise, blocking) raise the prediction residual in proportion to perceived degradation. The scalar $\mu$ enters only as a noise prior setting the absolute scale of this residual, explaining why so little reference information recovers most of the accuracy of a full-reference comparison on structural distortions.

The results reveal an interpretable dichotomy: PreSPA is highly competitive on structural distortions (blur, noise, compression, PIPAL restoration artefacts) but degrades on distortions that alter colour, hue, or contrast while leaving the gradient geometry intact, abundant in TID2013 and KADID-10K. This follows directly from analysing the luminance gradient field, and the fact that failures fall exactly where the structural assumption is violated confirms that the model behaves according to its principles. It also suggests a self-contained extension: a chromatic gradient channel would address colour distortions without altering the structural core.

Finally, the absence of learned weights yields a cross-dataset advantage. Deep models excel in-distribution but degrade under distribution shift, as seen in the drop of several of them on PIPAL, whereas PreSPA applies the same three fixed parameters to all six benchmarks with a markedly more stable ranking. Since its only data-dependent quantities ($\tau$ and $a$) are physically or statistically grounded rather than fitted on subjective scores, this predictability often outweighs a marginally higher in-distribution correlation when the distortion type is unknown a priori.

\section{Conclusion}
\label{sec:Conclusion}

\noindent We introduced PreSPA, a Partial-Reference IQA framework that reduces the reference signal to a single scalar prior $\mu$ and propagates it to an otherwise fully No-Reference analysis. Quality is decomposed into a structural index $Q_{\text{struct}}$, from a Hermite-Gauss self-prediction of the distorted gradient field interpreted as a normalised channel capacity, and a texture index $Q_t$ driven by $\mu$, which encodes the perceptual leakage of edge degradations into surrounding regions. Sharing a common perceptual scale through a viewing-distance-dependent kernel, the two indices are merged by a low-order affine fusion, yielding a model with three interpretable parameters and no learned weights.

Evaluated on six benchmarks, PreSPA consistently rivals or surpasses leading No-Reference deep models and approaches Full-Reference accuracy on structurally dominated distortions, at about $0.13$ GFLOPS per image pair. It inherits from BELE the edge--texture decomposition and the calibration-free treatment of viewing distance, being, to our knowledge, the first Partial-Reference method to embed the viewing distance into the operator scale, thereby reducing the VQEG logistic rectification to a plain affine alignment. Its training-free design also yields stable cross-dataset behaviour, valuable when the distortion is unknown a priori.

Operating on the luminance gradient field, PreSPA is intrinsically less sensitive to distortions that alter colour or contrast while preserving gradient geometry. Future work will add a chromatic gradient channel to cover such artefacts without altering the structural core, and will investigate adaptive estimation of $\mu$ in the absence of a reference, moving towards the fully No-Reference regime.

\bibliographystyle{IEEEtran}
\bibliography{IEEEabrv,imageprocessing}

\end{document}